\documentclass[journal]{IEEEtran}
\usepackage{etoolbox}
\usepackage{cite}
\usepackage{amsmath,amssymb,amsfonts}
\usepackage{algorithmic}
\usepackage{algorithm}
\usepackage{graphicx}
\usepackage{textcomp}
\usepackage{multirow}
\usepackage{booktabs}
\usepackage{bbding}
\usepackage{pifont}
\usepackage{adjustbox}
\usepackage{tabularray}
\usepackage{bm}
\usepackage{xcolor}

\usepackage{hyperref}
\hypersetup{
  colorlinks=true,
  linkcolor=blue,
  anchorcolor=blue,
  citecolor=blue
}

\newcommand{\yj}[1]{\textcolor{black}{#1}}

\begin{document}

\markboth{Journal, August~2025}%
{Shell \MakeLowercase{\textit{et al.}}: Bare Demo of IEEEtran.cls for IEEE Journals}

\title{Spatio-Temporal Representation Decoupling and Enhancement for Federated Instrument Segmentation in Surgical Videos}

\author{Zheng Fang, Xiaoming Qi, Chun-Mei Feng, \IEEEmembership{Member, IEEE}, Jialun Pei, \IEEEmembership{Member, IEEE}, \\ Weixin Si, \IEEEmembership{Senior Member, IEEE} and Yueming Jin, \IEEEmembership{Member, IEEE}
\thanks{Z. Fang, and Y. Jin are with the Department of Electrical and Computer Engineering, NUS, Singapore. Y. Jin and X. Qi are with the Department of Biomedical Engineering,
NUS, Singapore.}
\thanks{C. Feng is with the Institute of High Performance Computing, A*STAR, Singapore. J. Pei is with the Department of Computer Science and Engineering, The Chinese University of Hong Kong, HKSAR, China. W. Si is with Shenzhen Institutes of Advanced Technology, Chinese Academy of Sciences, Shenzhen, China.}
\thanks{Corresponding author: Yueming Jin (email: ymjin@nus.edu.sg)}}

\maketitle

\begin{abstract}
Surgical instrument segmentation under Federated Learning (FL) is a promising direction, which enables multiple surgical sites to collaboratively train the model without centralizing datasets. However, there exist very limited FL works in surgical data science, and FL methods for other modalities do not consider inherent characteristics in surgical domain: i) different scenarios show diverse anatomical backgrounds while highly similar instrument representation; ii) there exist surgical simulators which promote large-scale synthetic data generation with minimal efforts. In this paper, we propose a novel Personalized FL scheme, \textit{Spatio-Temporal Representation Decoupling and Enhancement} (FedST), which wisely leverages surgical domain knowledge during both local-site and global-server training to boost segmentation. Concretely, our model embraces a \textit{Representation Separation and Cooperation} (RSC) mechanism in local-site training, which decouples the query embedding layer to be trained privately, to encode respective backgrounds. Meanwhile, other parameters are optimized globally to capture the consistent representations of instruments, including the temporal layer to capture similar motion patterns. A textual-guided channel selection is further designed to highlight site-specific features, facilitating model adaptation to each site. Moreover, in global-server training, we propose Synthesis-based Explicit Representation Quantification (SERQ), which defines an explicit representation target based on synthetic data to synchronize the model convergence during fusion for improving model generalization. We construct a new PFL benchmark comprising five surgical sites from public datasets covering four types, with one out-of-federation site. FedST outperforms other state-of-the-art methods on federated sites (1.84\% on IoU) and achieves a remarkable improvement on the out-of-federation site (45.29\% on IoU).
\end{abstract}

\section{Introduction}
% segmentation
% \IEEEPARstart{S}{urgical} instrument segmentation in robot-assisted minimally invasive surgery (RAS) is critical \cite{maier2017surgical}, as it underpins advancements such as tool pose estimation \cite{pose_est}, precise tool tracking and control \cite{track}, and surgical task automation \cite{nagy2019dvrk}, significantly contributing to reduced trauma and accelerated patient recovery.   Intelligent parsing of such instruments, \cite{maier2017surgical, track, nagy2019dvrk}.
\IEEEPARstart{S}{urgical} instrument segmentation is fundamentally crucial to benefit robot-assisted minimally invasive surgery (RAS). Intelligent parsing of the instruments, including identifying their position and parts, not only provides cognitive assistance to surgeons \cite{loftus2020artificial,endovis2017,endovis2018}, but also lays the foundation for higher-level downstream tasks such as instrument pose estimation \cite{pose_est}, tracking \cite{track}, and task automation \cite{nagy2019dvrk}.
% part seg
Although the deep learning model has shown promise in surgical instrument segmentation \cite{jin2022exploring, kondo2021lapformer, wang2021noisy, liu1}, as a data-driven approach, it heavily relies on the data quantities to facilitate its efficacy \cite{he2016deep}. Collaborative training using different RAS datasets from multiple clinical sites could benefit from maximizing its potential in surgical instrument segmentation. 
%%but limited by quantity and quality of the available data \cite{he2016deep}.
 % collaborative in current methods
%Collaborative training on multi-center datasets from diverse clinical sites enhances the advancement and generalizability of surgical instrument segmentation models.
% 
Direct data communication over multiple sites is however infeasible due to privacy protection for patients. Federated Learning (FL) has been an important topic, enabling practical collaborative training without data sharing and privacy concerns \cite{kaissis2020secure}.
%第一段把surgical instrument segmentation的意义，以及collaborative training对于提升instrument segmentation的价值写出来，可以参考https://arxiv.org/abs/2408.03208

%% Related Work
% FL and PFL
%FL has limited application in surgical scenarios, despite their widespread use in other medical fields.
% FL
General FL methods train a global model by sharing model parameters and perform parameter aggregation on the server side \cite{fedavg}. %which has proven effective in varying medical segmentation tasks\cite{li2019privacy, dou2021federated}. 
However, with a single model, this method struggles to perform well for all sites given data heterogeneity, especially in surgical scenarios, arising from different procedures, surgical protocols, equipments, and patient characteristics. 
%primarily due to the used server-side average aggregation strategy, such as FedAVG\cite{yu2022salvaging}.
% \cite{yu2022salvaging, wang2019federated}, 
% PFL
Personalized Federated Learning (PFL) has been proposed to meet such limitation by generating multiple local site models. PFL methods generally divide model parameters into global and personalized parts, with personalized ones commonly including prediction heads \cite{fedrep}, batch normalization \cite{li2021fedbn}, convolution
channels \cite{shen2022cd2}, or query embeddings in self-attention \cite{feddp}. Another popular stream of PFL approaches focuses on global server optimization, such as assigning different weights to each site in global aggregation, by using the Fisher Information Matrix \cite{yang2024fedas} or decomposing uploaded gradients \cite{zeng2024tackling}.
%PFL的related work感觉写的不充足，至少类别应该涵盖了我们所有的对比方法，你看看我的几篇TMI paper。这里可以把怎么去separate parameter，以及哪几种partial parameter都写出来
These methods have demonstrated success in varying medical image segmentation tasks, such as polyp segmentation \cite{feddp} and optic disc/cup segmentation \cite{jiang2023iop}. %where personalization is crucial for handling domain shifts and uncertainties 
% \cite{li2021ditto, jiang2023iop}
% Recent state-of-the-art approaches such as FedLD \cite{zeng2024tackling}, which performs gradient decomposition on the server side before aggregation, and FedDP \cite{feddp}, which privatizes the query layer parameters, have achieved impressive performance in medical image segmentation. 
However, \emph{PFL for instrument segmentation from surgical videos still remains under-explored.} The most related work in the surgical domain is for workflow recognition \cite{fedcy}, which is an image-level classification task and implements FL across sites with the same surgical type, i.e., cholecystectomy. There remains a large research gap on how to perform FL on pixel-level segmentation tasks across different surgical types, for better clinical applicability.
%
% Federated Learning (FL) has been widely applied for collaborative training in various medical scenarios. The FL is proposed to address the challenges posed by the isolated data island and privacy preservation issues in deep learning methods\cite{fedavg}. In \cite{li2019privacy}, FedAvg is directly applied to data from different centers to collaboratively train a brain tumor segmentation model. \cite{dou2021federated} constructs a dataset of lung CT scans from multinational hospitals to detect COVID-19 anomalies and applies semi-supervised learning to enhance model training. However, the data heterogeneity, especially the different medical scanning devices and, protocols, limits the application of FL in real-world medical scenarios\cite{wang2019federated, yu2022salvaging}. To overcome this, Personalized Federated Learning (PFL), a mainstream branch of FL, is proposed to enhance the performance of local site models\cite{pfl}. The existing PFL works attempt decouple partial model parameters for local optimization \cite{chen2021personalized, fedrep, li2021fedbn} and distill knowledge for the the adaptation on different distributions \cite{li2021ditto,jiang2023iop}. These techniques also have been applied to medical application to tackle challenges specific to data heterogeneity, domain shift, uncertainty\cite{liu2021feddg, FedMix2022}.%
%% Challenge 

\begin{figure}[t]
\centering
\includegraphics[width = 0.5\textwidth]{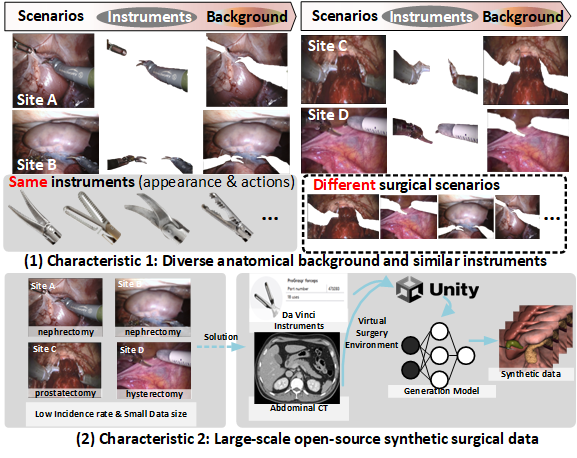}
\caption{(1) Surgical procedures from four centres covering three types, show diverse anatomical backgrounds, while instruments exhibit high visual and motion consistency. (2) Simulators can generate large-scale data with minimal effort, while the domain gap between synthesis and real data needs to be well tackled for effective data augmentation.} \label{challenge}
\vspace{-2mm}
\end{figure}

Existing PFL methods also have limitations with suboptimal results when adapting to the surgical scenario, without considering the natural characteristics of surgical procedure. We have identified two of them:
\textbf{\textit{(i) Diverse anatomical background and similar instruments}}. Background tissue in different types of surgery and hospitals exhibits a large diversity. %making it challenging for local site models to establish a consistent representation of surgical instruments.
Instead, there exists a high similarity between robotic instruments from different scenarios, in terms of both \textit{spatial appearance} and \textit{temporal motion} patterns \cite{jin2022exploring} (cf. Fig.\ref{challenge} (1)). 
Specifically, surgical instruments, especially the surgeries using Da Vinci robots, generally show similar shapes and consist of shaft, wrist, and jaw parts. 
Additionally, compared to irregular deformation of the background tissues, instrument motion is more uniform and pronounced across different surgical scenarios.
% example
% For instance, instruments like Maryland Bipolar Forceps or Prograsp Forceps maintain similar shapes and perform analogous tissue-cutting actions, which can be leveraged to achieve consistent modeling across diverse scenarios. 
For instance, instruments like Maryland Bipolar Forceps or Prograsp Forceps present similar shapes and commonly perform analogous cutting action. Such consistency property of instruments can be leveraged in collaborative segmentation learning even across different surgical types. 
%
%Therefore, we aim to personalize the model to adapt to diverse backgrounds while learning a globally consistent representation of the instruments, encompassing both visual appearance and temporal motion. 

\textbf{\textit{(ii) Large-scale open-source synthetic surgical data}}. Compared to other medical modalities, there exist simulators in the surgical domain, which can generate simulated surgical data with corresponding ground truths (cf. Fig.\ref{challenge} (2)). This process requires no manual annotation from clinical experts, therefore facilitating data generation with minimal effort and yielding large-scale simulated data available as open-source \cite{siseve}. It can be readily utilized to benefit global server training in FL. However, despite some studies employing style transfer techniques to synthesize the simulated data to appear more realistic, the substantial domain gap from real surgical videos cannot be overlooked. How to effectively leverage such synthesis data in FL needs further investigation.
In this study, we propose a novel PFL scheme, \textbf{S}patio-\textbf{T}emporal Representation Decoupling and Enhancement  (FedST) for federated surgical instrument segmentation, to achieve collaborative learning across various surgical types.
%
% To avoid the influence of background diversity in FL, we propose a Representation Separation and Contrastive Cooperation (RSC) to introduce an instrument consistent representation target across various scenarios from the similar appearance and temporal motion of instruments. First, RSC separates the instrument's consistent representation by spatial-temporal modeling on previous frames and stores the local unique representation by privatizing partial parameters. Then, RSC identifies the difference between the global fusion and the site's representation by integrating the pre-trained CLIP as the contrastive adaptation to enhance the representation of each site. RSC not only ensures that the FL aggregation phase is guided by a globally consistent objective but also ensures the adaptation of FL aggregation on each site's representation.
% One sentence for RSC
% We aim to personalize the local model to adapt to diverse backgrounds while learning a globally consistent representation of the instruments, encompassing both visual appearance and temporal motion. 
Our method aims to effectively incorporate the aforementioned surgical domain knowledge into the two FL phases, i.e., local-site and global-server learning, to boost segmentation. 
Concretely, we first propose a Representation Separation and Cooperation (RSC) mechanism to facilitate local training. It personalizes the query embedding layers to decouple the specialized background representation of each local site. While other parameters are optimized cooperatively to encode the spatio-temporal representation of instruments given its consistency across different sites. 
Additionally, we introduce the textual description for each scenario as an indicator to emphasize site-specific representation, further facilitating personalized adaptation to each site.
% It defines the partial sharing of key and value in attention-based module to decouple the spatio-temporal representation of the instrument from various scenarios for generalization enhancement and then takes the text-based description of each scenario as an indicator to cooperate with the enhanced instrument representation to each scenario.
%personalize the local model to adapt to diverse backgrounds while learning a globally consistent representation of the instruments, encompassing both visual appearance and temporal motion. This module leverages multi-head cross-attention to capture temporal motion patterns of instruments while decoupling the background tissue representation by personalizing the query parameters. Additionally, CLIP-based text guidance is used as an indicator, along with a lightweight network, to guide and enhance the representation of the instruments. 
Moreover, we propose a Synthesis-based Explicit Representation Quantification (SERQ) strategy to benefit global-server learning, by effectively exploiting synthetic data. 
% On the server side, SERQ explicitly quantifies a learnable domain descriptor to separate the instrument representation from synthetic data and synchronizes the instrument representations from synthetic data  various sites. 
SERQ explicitly quantifies a learnable domain descriptor to separate instrument representation from synthetic data, which is then used to enhance the global model and synchronize the convergence of local models. The learnable domain descriptor and synchronization strategy can reduce the domain gap between synthetic data and local site data, to enhance representation learning without bringing any extra cost during inference. Our resource code will be available. 

Our main contributions can be summarized as follows: 
\begin{enumerate}
    \item We introduce a novel PFL scheme (FedST) for instrument segmentation from surgical videos, designed to identify cross-site consistent representations and synchronize the convergence of multi-site optimization by leveraging the synthetic dataset. 
    % Our approach provides a more practical and privacy-preserving PFL solution adaptive to different surgical scenarios and boosts the performance of downstream segmentation tasks.
    \item We propose the RSC mechanism to construct consistent spatio-temporal representations of instruments at local sites while personalizing site-specific variations across diverse tissue backgrounds.
    % to maintain the consistent instrument representation in FL from different scenarios with various tissue backgrounds. 
    % It leverages motion-induced spatial-temporal modeling and textual-guided channel selection to learn a globally consistent representation of instruments.
    \item We design the SERQ strategy to effectively leverage instrument representations from the synthetic dataset, enhancing the global model and synchronizing local model convergence. 
    % It defines an explicit quantification target for instrument consistent representation and synchronizes model convergence by transferring the convergence standard from pre-training on a large synthetic dataset.
    %%SERQ module is proposed to tackle limited data availability across multi-hospital sites. SERQ defines an explicit quantification target for instrument consistent representation and synchronizes model convergence by transferring the convergence standard from pre-training on a large synthetic dataset.  
    \item We construct a new benchmark for federated instrument segmentation from surgical videos, consisting of five surgical sites, one being out-of-federation, and covering four different surgical types. Experiments have demonstrated the superior performance of our method compared to state-of-the-art methods, especially in the out-of-federation site.

\end{enumerate}

\section{Related Work}
\subsection{Surgical Instrumentation Segmentation}
Medical image segmentation plays a crucial role in clinical settings within hospitals~\cite{wang1,wang2,wang3,wang4}, serving as a foundation for automated disease diagnosis~\cite{fu1,fu2,fu3} and precise treatment planning~\cite{yang2025medical}.
Extensive studies have conducted on surgical instrument segmentation. Techniques such as holistically-nested networks \cite{garcia2017toolnet}, graph-based networks \cite{liu2021graph}, synergistic networks \cite{wang2024video}, contrastive learning \cite{lou2023min}, activate learning \cite{peng2024reducing}, and utilizing auxiliary cues like depth maps \cite{mohammed2019streoscennet}, optical flow \cite{jin2019incorporating}, motion flow \cite{zhao2020learning}, tracking cues \cite{zhao2022trasetr}, and synthetic images \cite{colleoni2021robotic} have been investigated. Recently, the advent of the Segment Anything \cite{kirillov2023segment} model has heralded a new era for segmentation tasks. Fine-tuning SAM has been explored to bridge the domain gap between surgical and natural images \cite{surgsam,liu2024surgical}, with SurgSAM \cite{surgsam}, for example, incorporating a trainable prototype-based class prompt encoder to enhance segmentation accuracy. However, most existing methods for surgical instrument segmentation focus on training models locally within one surgical type without taking advantage of collaborative training to improve the model performance.

\subsection{Federated Learning}
Owing to data privacy concerns, FL has attracted increasing attention in fields, especially in the medical domain, as it enables decentralized model training across sites without sharing raw data and has achieved notable success \cite{fedavg,silva2019federated,guan2024federated,subedi2023client}.
FL has been applied to brain tumor segmentation by using FedAvg to combine insights from different centers \cite{li2019privacy}. It has also been utilized for detecting COVID-19 anomalies in lung CT scans, incorporating semi-supervised learning to boost performance \cite{dou2021federated}. FL faces limitations due to data heterogeneity, such as differing medical imaging devices and protocols \cite{yu2022salvaging}.

To mitigate these issues, PFL has been proposed to allow partial model parameter personalization for each local site, so that the federated model can be adapted to distinct data distribution across different sites. PFL divides model parameters into global and personalized layers, such as prediction heads \cite{fedrep}, batch normalization \cite{li2021fedbn}, convolution channels \cite{shen2022cd2}, or query embeddings in self-attention \cite{feddp}. Additionally, server-side optimization approaches, such as weighting sites using the Fisher Information Matrix \cite{yang2024fedas} and decomposing uploaded gradients \cite{zeng2024tackling}, have been explored to further enhance PFL's adaptability to heterogeneous data.
Despite advancements in medical applications, investigations into applying FL and PFL for surgical instrument segmentation remain limited. Existing FL methods often overlook key characteristics of the surgical domain, such as the intricate temporal nature of video data, resulting in suboptimal performance and underscoring the need for further research in this area.

\begin{figure*}
  \includegraphics[width=\textwidth]{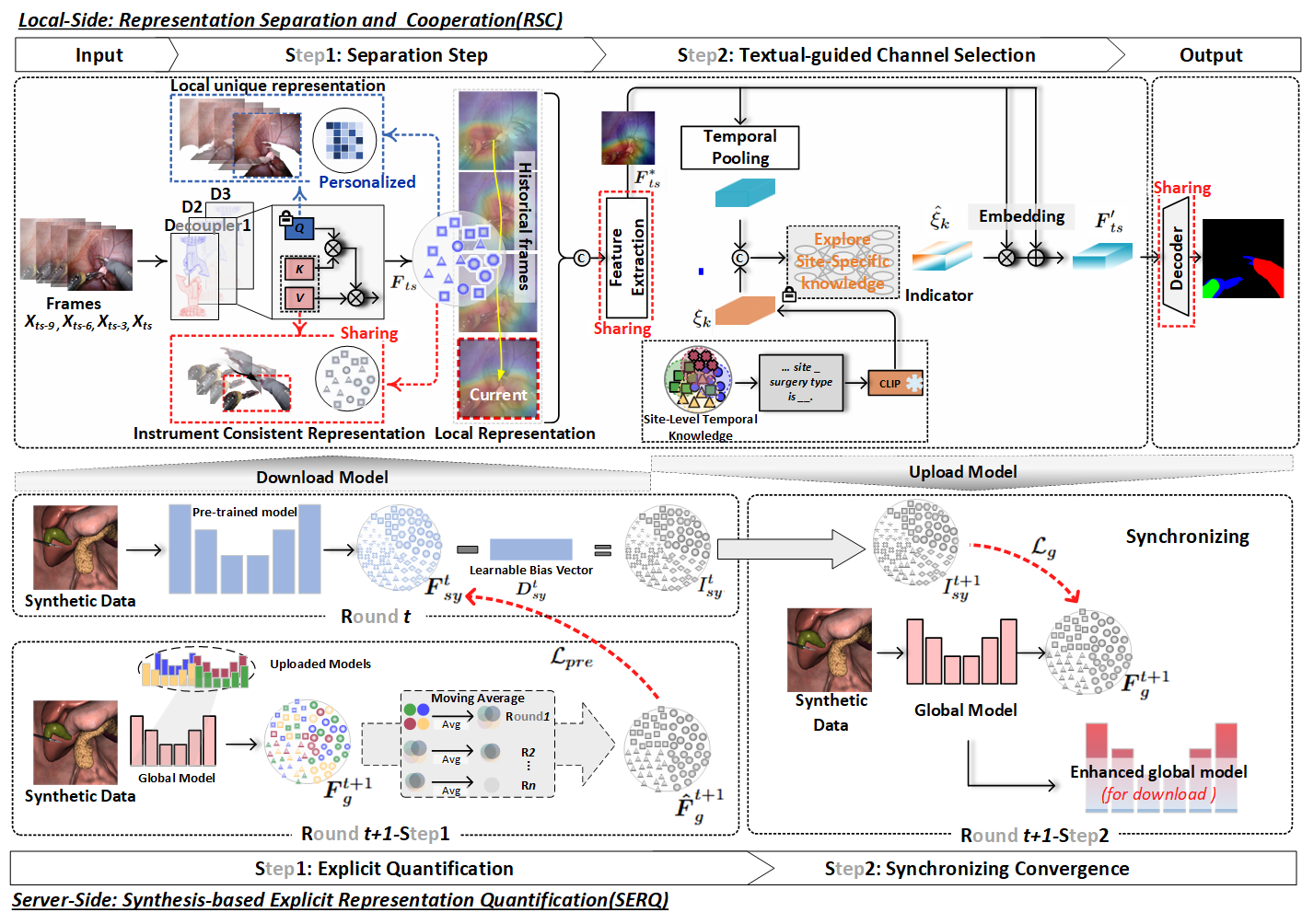}
  \vspace{-0.6cm}
  \caption{Overall pipeline of our proposed FedST. RSC separates the instrument consistent representation as the target of FL optimization by motion-induced modeling and textual-guided channel selection. SERQ explicitly quantifies the instrument consistent representation and enhances the global model by synthesis dataset.} 
  \label{method}
\end{figure*}
\section{Methodology}

\subsection{Overall Pipeline}
\subsubsection{Problem Formulation}
Assume that there are $K$ local sites, each with a unique non-iid distribution $\mathcal{D}_k$ over a set of $N_k$ samples $\left\{\left(x_{i,k}, y_{i,k}\right)\right\}_{i=1}^{N_k}$, where $\left(x_{i,k}, y_{i,k}\right) \in \mathcal{D}_k$. The joint image and label space across these sites is denoted as $\left\{\left(\mathcal{X}_k, \mathcal{Y}_k\right)\right\}_{k=1}^K$.
The objective of the general FL framework, specifically FedAvg \cite{fedavg}, is to learn a global model by minimizing the expected loss across all sites. This can be formalized as follows:
\begin{equation}
\label{gfl}
w^{t+1} = w^t - \eta \mathbb{E}_{k \in[K]} \mathbb{E}_{\left(x_{i, k}, y_{i, k}\right) \in \mathcal{D}_k} \nabla \mathcal{L}\left(f_w\left(x_{i, k}\right), y_{i, k}\right),
\end{equation}
where $w^t$ represents the model parameters at round $t$; $\eta$ is the learning rate; and $\mathcal{L}$ is the loss function. PFL methods propose to learn the respective local models $\{v_k\}_{k=1}^K$ for all the local sites, where each local model $v_k$ aims to suit the $k$-th site's feature distribution thus we reformulate Eq. \ref{gfl} as:
\begin{equation}
\label{pfl}
v_k^{t+1} = v_k^t - \eta \mathbb{E}_{(x_{i,k}, y_{i,k}) \in \mathcal{D}_k} \nabla \mathcal{L}(f_{v_k}(x_{i,k}), y_{i,k}).
\end{equation}
% We focus on personalizing the query embedding layers in the transformer-based encoder while sharing the remaining parameters during the FL process, enabling adaptation to each site's unique tissue background.
 
% Optimizing solely for the global model often overlooks the specific data distribution in each local model. PFL decouples the parameters of each local model into $\theta_k = (\gamma_k, \eta_g)$, where $\gamma_k$ represents the local parameters that do not participate in the aggregation to global server, and $\eta_g$ are the parameters that are involved in the model aggregation. Therefore, we reformulate the objective function in Eq. \ref{obj1} as follows:
% \begin{equation}
%     (\eta_{g}^{*}, \gamma_{k}^{*}) = \argmin_{\eta_g, \{\gamma_k\}_{k=1}^{K}} \left[ \sum_{k=1}^{K} k_{m}\mathcal{L}_k(\eta_g, \gamma_k ; \mathcal{D}_k) \right].
%     \label{obj2}
% \end{equation}

\subsubsection{Overview of FedST}
Our FedST provides a PFL framework to achieve surgical instrument segmentation in a decentralized manner. It consists of Representation Separation and Cooperation (refer to RSC Sec. \ref{RSC}) and Synthesis-based Explicit Representation Quantification (refer to SERQ Sec. \ref{SERQ}). The overall pipeline is shown in Algorithm \ref{alg:frame}. 
FedST operates in two stages: In the local-side stage, RSC is performed to personalize the local model to adapt to diverse backgrounds by focusing on temporal dynamics and site-specific background adaptation. In the server-side stage, SERQ extracts the instrument representation from synthetic data to enhance the global model and synchronize the convergence of local site models. 
Regarding the training procedure, we first use the synthetic dataset for pre-training to initialize local site models.
\begin{figure}[!t]
		\renewcommand{\algorithmicrequire}{\textbf{Input:}}
		\renewcommand{\algorithmicensure}{\textbf{Output:}}
		\begin{algorithm}[H]
			\caption{Federated Learning via Spatial-Temporal Decoupling} \label{alg:frame}
			\begin{algorithmic}[1]
				\REQUIRE Local Datasets $\left\{\left(\mathcal{X}_k, \mathcal{Y}_k\right)\right\}_{k=1}^K$, Synthetic Dataset $\left\{\left(\mathcal{X}_{sy}, \mathcal{Y}_{sy}\right)\right\}$, Pre-trained Model $v_{pre}^{0}$, Number of communication rounds $T$.
				\ENSURE Personalized local models $\left\{v_k\right\}_{k=1}^K$. %%output
				\STATE Initialize personalized parameters for each $ \left\{\rho_k\right\}_{k=1}^K$.
                \STATE Initialize other shared parameters for each $ \left\{\gamma_k\right\}_{k=1}^K$.
                \STATE Initialize global model $ v_{g}^{0} \leftarrow v_{pre}^{0} $.
				\FOR {$t = 0,1, ... , T - 1 $}
                    \FOR{$k = 0,1, ... , K - 1 $}
                        % \STATE $v_k^{t+1} :=  \gamma_g^{t} \cup \rho_k^{t} $
                        % \STATE $\rho_k^{t},\gamma_k^{t} \leftarrow RSC \left\( \left(\mathcal{X}_k, \mathcal{Y}_k\right\), \gamma^{t},\rho_k^{t} \right\) $
                        \STATE $\rho_k^{t+1}, \gamma_k^{t+1} \leftarrow \textcolor{red}{\textbf{RSC}}\left(\left(\mathcal{X}_k, \mathcal{Y}_k\right), \rho_k^{t},\gamma_g^{t}\right)$
                        
                    \ENDFOR
                    \STATE $\gamma^{t+1}_g := \frac{1}{K} \sum_{k=1}^{K} \gamma_k^{t+1}$.
                    \STATE $v_g^{t+1} :=  \gamma_g^{t+1} \cup \rho_g^{t} $

                    \textcolor{blue}{/* Extract representation in SERQ */} \hfill
                    
                    \STATE $v_{sy}^{t+1} \leftarrow \textcolor{red}{\textbf{SERQ}_{EQ}}\left(\left(\mathcal{X}_{sy}, \mathcal{Y}_{sy}\right), v_{pre}^{t}, v_g^{t+1}\right)$
                    
                    % \STATE Update Pre-trained model $v_{pre,t}$ via Eq.~(\ref{fg}) and (\ref{mse}). 

                    \textcolor{blue}{/* Synchronize the convergence in SERQ */} \hfill
                    
                    \STATE $v_{g}^{t+1} \leftarrow \textcolor{red}{\textbf{SERQ}_{SC}}\left(\left(\mathcal{X}_{sy}, \mathcal{Y}_{sy}\right), v_{pre}^{t+1}, v_g^{t+1}\right)$
                    
                    \STATE Decouple $v_{g}^{t+1}$ and distribute $\gamma_{g}^{t+1}$ to local sites.
                    
                \ENDFOR
                
                \textcolor{blue}{/* Merge Parameters */} \hfill

                \FOR{$k = 0,1, ... , K - 1 $}
                    \STATE $v_k :=  \gamma_g^{T} \cup \rho_k^{T} $
                \ENDFOR
				\STATE Return personalized local models $\left\{v_k\right\}_{k=1}^K$.
			\end{algorithmic}
		\end{algorithm}
	\end{figure}
Then we perform $T$ rounds of local-global communication with $E$ local updates per round. The local models are optimized by using RSC and the server performs SERQ every round by receiving the updated parameters of all sites. We can obtain personalized models after $T$ rounds of communication without disclosing any local private data. 

% To facilitate the decoupling of parameters later, we divide the local model parameters into query parameters $\gamma_k$ for local unique representation and other parameters $\eta_k$ for instrument consistent representation:
% $
% v_k = (\gamma_k, \eta_k)
% $
% , where $\gamma_k$ represents the query parameters and $\eta_k$ represents the other parameters. The query parameters $\gamma_k$ have no intersection with the other parameters $\eta_k$: 
% $\gamma_k \cap \eta_k = \emptyset , \forall k \in \{1, 2, \ldots, K\}$.
% Therefore, the set of personalized local models can be expressed as:
% $\left\{v_k\right\}_{k=1}^K = \left\{(\gamma_k, \eta_k)\right\}_{k=1}^K, \text{where } \gamma_k \cap \eta_k = \emptyset \quad \forall k \in \{1, 2, \ldots, K\}
% $. 

\subsection{Representation Separation and Cooperation}
\label{RSC}
% RSC employs a two-pronged approach: (1) Separation step. It separates and stores the local unique representation and instrument consistent representation in the model parameters based on the appearance and temporal motion. (2) Textual guided channel selection. It utilizes the pre-trained CLIP to cooperate with the representation of each site, ensuring the adaptation to each site. 
RSC separates the local unique representation and the global consistent representation by determining which portions of the model parameters should be personalized or shared, so that the consistent representation of instruments can be effectively cooperated across sites. Additionally, a textual-guided channel selection leverages pre-trained CLIP to further enhance the channels relevant to local site features, ensuring better adaptation to diverse surgical scenarios.
\subsubsection{Separation and Cooperation of Visual Representation}
% Motivation ： 这里的第一段需要从PFL去写，也就是说明下 
%1因为xxx, 所以分成personal&glocal parameter会好；
%2 instrument因为appearance&motion across site相似 所以做global；tissue因为xx 做personal
%我们为了更好的capture instrument feature，做了xxx

PFL typically decouples partial model parameters for local updates, allowing models to adapt to site-specific distributions. This strategy is effective in handling data heterogeneity by combining global and personalized knowledge. In RAS, instruments exhibit consistent appearance and motion patterns across sites, making them suitable for global parameter sharing. Conversely, tissue representations vary significantly across sites and require personalized parameterization to capture site-specific background variations effectively.
Thus, we aim to capture consistent instrument motion across sites by incorporating both the current and previous frames, using a multi-scale approach with larger receptive fields for distant frames, as shown in Fig. \ref{method}.
% \cite{sun2022coarse,wang2022pvt,yang2021focal}
% Simple workflow and why it will work

Specifically, given training data containing video frames $\left\{\boldsymbol{x}_{ts-m}, \cdots, \boldsymbol{x}_{ts-1}, \boldsymbol{x}_{ts}\right\}$ with ground truth of segmentations $\left\{\boldsymbol{y}_{ts-m}, \cdots, \boldsymbol{y}_{ts-1}, \boldsymbol{y}_{ts}\right\}$, we aim to segment current frame $\boldsymbol{x}_{ts}$.
We first utilize a self-attention encoder, which includes partially personalized parameters, to extract visual features from the video clip, denoting $\left\{\boldsymbol{F}_{ts-m}, \cdots, \boldsymbol{F}_{ts-1}, \boldsymbol{F}_{ts}\right\}$, where $\boldsymbol{F} \in \mathbb{R}^{h \times w \times c}$, with $h$, $w$, and $c$ representing height, width, and the number of channels. The details of these personalized parameters will be discussed later.
We then split the current frame feature $\boldsymbol{F}_{ts}$ into windows of size $s \times s$ to reduce computational cost, where each window attends to a set of context tokens: $\boldsymbol{F}_{ts} \in \mathbb{R}^{h \times w \times c} \rightarrow \boldsymbol{F}_{ts} \in \mathbb{R}^{\frac{h}{s} \times \frac{w}{s} \times s \times s \times c}.$
% \begin{equation}
%     \label{t_split}
%     \boldsymbol{F}_t \in \mathbb{R}^{h \times w \times c} \rightarrow \boldsymbol{F}_t \in \mathbb{R}^{\left(\frac{h}{s} \times s\right) \times \left(\frac{w}{s} \times s\right) \times c} \rightarrow \boldsymbol{F}_t \in \mathbb{R}^{\frac{h}{s} \times \frac{w}{s} \times s \times s \times c}.
% \end{equation}
%\begin{equation}
%    \label{t_split}
%    \boldsymbol{F}_{ts} \in \mathbb{R}^{h \times w \times c} \rightarrow \boldsymbol{F}_{ts} \in \mathbb{R}^{\frac{h}{s} \times \frac{w}{s} \times s \times s \times c}.
%\end{equation}

Considering that instruments are typically moving while tissue backgrounds remain relatively static, we partition the current and previous features using different pooling sizes $p$ and different receptive field sizes $r$ to identify all possible positions of the instruments. Partitioned features are subsequently processed by fully connected layers (FC) for refinement:
\begin{equation}
    \label{j_split}
    \boldsymbol{F}_{ts-m} \in \mathbb{R}^{\frac{h}{p} \times \frac{w}{p} \times c \times p^2} \xrightarrow{\mathrm{FC}} \boldsymbol{F}_{ts-m} \in \mathbb{R}^{\frac{h}{p} \times \frac{w}{p} \times c}.
\end{equation}
% \boldsymbol{F}_{ts-m} \in \mathbb{R}^{h \times w \times c} \rightarrow \boldsymbol{F}_{ts-m} \in \mathbb{R}^{\frac{h}{p} \times \frac{w}{p} \times c \times p^2} \xrightarrow{\mathrm{FC}} \boldsymbol{F}_{ts-m} \in \mathbb{R}^{\frac{h}{p} \times \frac{w}{p} \times c},
%
For each window partition in the current frame, we perform multi-head cross-attention to capture dependencies between the current frame and previous frames, constructing a temporal representation for each partition. We first define $\boldsymbol{c} \in \mathbb{R}^{\frac{h}{s} \times \frac{w}{s}}$ to represent the context tokens for each window partition in the current features $\boldsymbol{F}_{ts}$. We then extract elements from each window partition in $\boldsymbol{F}_{ts-m}$ and concatenate them into $\boldsymbol{c} $, creating a unified context representation $\boldsymbol{\hat{c}}$ for each partition in $\boldsymbol{F}_{ts}$. 
The augmented context tokens $\boldsymbol{\hat{c}}$ encompass possible positions of the instrument in previous frames. Consequently, we use multi-head cross attention to construct the feature map $\boldsymbol{F}_{ts}^*$ of the current frame, which incorporates the temporal representation. For simplicity, multi-head cross attention for each window partition in the current frame is expressed as:
\begin{equation}
    \label{multi}
\boldsymbol{F}_{ts}^*=\operatorname{Softmax}\left(\frac{Q K^T}{\sqrt{c}}+B\right) V+\boldsymbol{F}_{ts},
\end{equation}
where $B$ is the position bias; c is the number of channels; and $Q = \mathrm{FC}(\boldsymbol{F}_{ts})$, $K = \mathrm{FC}(\boldsymbol{\hat{c}})$, and $V = \mathrm{FC}(\boldsymbol{\hat{c}})$ are the query, key, and value vectors obtained from FC operations.

We have constructed the temporal representation of the instrument, which remains consistent across different sites. However, to account for the site-specific variations in background tissue representation, which lacks temporal information and differs significantly across sites, we personalize these representations by privatizing the query embedding parameters $\rho$ in the self-attention-based encoder. Because the query embedding layers capture the unique details of each pixel \cite{feddp}, allowing each local site to capture site-specific background tissue representations more effectively. Let $v$ represent the entire parameter set of the local site model. We decouple $v$ into two parts: $\rho$ denote the parameters of the query embedding layers in the attention-based encoder, and $\gamma$ denote the remaining parameters that are shared across sites. Formally, this decoupling can be expressed as $v \rightarrow \rho \cup \gamma$ with $\rho \cap \gamma = \emptyset$. 
To update the model, we separate the parameters into private and global parts. The private parameters $\rho_k$ in each site $k$ are updated locally using gradient descent:
\begin{equation}
    \label{p_param}
    \rho_k^{t+1} = \rho_k^t - \eta \mathbb{E}_{(x_{i,k}, y_{i,k}) \in \mathcal{D}_k} \nabla \mathcal{L}(f_{v_k}(x_{i,k}), y_{i,k}),
\end{equation}
and the shared global parameters $\gamma$ are aggregated from all local models through an averaging approach, using global gradients computed across all sites:
\begin{equation}
    \label{other_param}
    \gamma^{t+1} = \gamma^t - \eta \mathbb{E}_{k \in [K]} \mathbb{E}_{(x_{i,k}, y_{i,k}) \in \mathcal{D}_k} \nabla \mathcal{L}(f_{v_k}(x_{i,k}), y_{i,k}).
\end{equation}
This dual-update strategy allows the local model to maintain a globally consistent representation of the instruments, while also capturing site-specific background representation.

\subsubsection{Textual-guided Channel Selection}
Although instrument representations are consistent across sites and shared globally, each site still retains distinctive characteristics due to variations in surgical types and utilized robots. \yj{Therefore, we incorporate site-specific uniqueness into the local model’s representation learning to better adapt to and enhance the feature representations specific to each site.} We propose a textual-guided channel selection approach (shown in Fig \ref{method}) using a personalized lightweight network and a pre-trained CLIP model with pre-defined text that captures site-level knowledge to separate the site-specific representations across different sites. This site-level knowledge represents the uniqueness of each site's instrument representation. 

Formally, we use pre-defined text \textit{This is the Local site X model and the surgery type is XXX} as input text prompts for the pre-trained {frozen} CLIP model, generating an indicator \(\xi_{k} \in \mathbb{R}^{h \times w}\) for the $k$-th local site model. 
This indicator is designed to assist the local model in channel selection, enabling it to focus on relevant features within each site’s data. \yj{Since the CLIP text encoder is frozen, the semantic embedding $\xi_k$ remains fixed during training, allowing each site to consistently enhance the global representation along a stable, site-specific semantic direction.}
To assist the indicator \(\xi_k\), we use the global average pooled values derived from \(\boldsymbol{F}_{ts}^* \in \mathbb{R}^{h \times w \times c}\) and concatenate it with the indicator $\xi_k$ and passed through a fully connected layer followed by a sigmoid activation $\sigma$ to produce the final composite indicator, denoted as $\hat{\xi}_k \in \mathbb{R}^{h \times w}$. This transformation is represented as follows:
\begin{equation}
    \label{hat}
    \hat{\xi}_k = \sigma(\mathrm{FC}(\text{Concat}(\text{AvgPool}(\boldsymbol{F}_{ts}^*), \xi_k))).
\end{equation}
The enhancement is performed by integrating the composite indicator $\hat{\xi}_k$ with the feature map $\boldsymbol{F}_{ts}^*$ through pixel-wise multiplication and a residual operation in case of wrong channel selection, as shown below:
\begin{equation}
    \label{trc}
    \boldsymbol{F}_{ts}' = \boldsymbol{F}_{ts}^* + \boldsymbol{F}_{ts}^* \otimes \hat{\xi}_k,
\end{equation}
where $\otimes$ denotes pixel-wise multiplication. This operation selectively enhances channels that are aligned with the indicator. The resulting $\boldsymbol{F}_{ts}'$ represents the feature map after enhancement, with an improved focus on relevant information.
Finally, the enhanced features \(\boldsymbol{F}_{ts}'\) are fed into the decoder to generate the segmentation mask. 

In this local-side training phase, we employ a combination of Dice loss and Cross-Entropy loss as the objective function:  
\begin{equation}
\label{total}
\mathcal{L}_{\text{Seg}} = \mathcal{L}_{\text{CE}} + \lambda_1 \mathcal{L}_{\text{Dice}},
\end{equation}
where $\lambda_1$ is introduced to control the contribution of the Dice loss in the overall segmentation loss.

% where the Dice loss $\mathcal{L}_{\text{Dice}}$ is defined as:
% \begin{equation}
% \label{dice}
% \mathcal{L}_{\text{Dice}} = 1 - \frac{2 \sum_{i=1}^N \mathcal{Y}_k(i) P_k(i)}{\sum_{i=1}^N \mathcal{Y}_k(i) + \sum_{i=1}^N P_k(i)},
% \end{equation}

% synthetic
\subsection{Explicit Representation Quantification}
\label{SERQ}  
% For the server-side training, SERQ defines a learnable domain descriptor to separate the instrument representation from synthetic data and synchronizes the instrument representations from synthetic data and various sites to enhance the model generalization in FL. The learnable domain descriptor and synchronize strategy eliminate the domain gap among synthetic data and various scenarios and enhance the instrument consistent representation.
%We have achieved consistent instrument representation on the local-side; however, due to varying data volumes across sites, the convergence levels of local models differ, potentially leading to insufficient extraction of instrument representations.
Apart from benefiting adaptive representation learning in local-site training, we aim to further enhance the feature representation during global server-side training by leveraging large-scale open-source synthetic data. 
We propose SERQ which defines a learnable domain descriptor to extract the instrument representation from synthetic data and synchronize it with the instrument representations from various sites, thereby enhancing model generalization in FL. The learnable domain descriptor and synchronization strategy effectively reduce the domain gap between synthetic data and local site data while improving the consistency of instrument representations.

%we first designed to explicitly quantify the instrument representation of the synthetic dataset, which was achieved by using the moving average of the global model features. The quantified instrument representation is then utilized to enhance the global model to synchronize the convergence of the different local site models.
% To prevent the global model from being biased towards data-rich sites due to limited data availability, we propose SERQ, which leverages synthetic data free of privacy concerns. First, we use a moving average of the global model features to iteratively update a pre-trained model on the synthetic dataset, explicitly quantifying the instrument representation and separating it from domain-specific knowledge, such as style or background representation. This instrument representation is then utilized to synchronize the convergence of the global model, enhancing the global model's ability to capture the instrument-consistent representation.

\subsubsection{Explicit Quantification} 
% Directly incorporating synthetic data can harm the global model due to the significant domain gap between synthetic and real datasets. Inspired by \cite{zhang2024eliminating}, which decomposes features into domain-specific and general components, we define the representation \( \boldsymbol{F} \) extracted from the encoder as the combination of do main representation and instrument-consistent representation: \( \boldsymbol{F} = \boldsymbol{F}^D + \boldsymbol{F}^I \). Since synthetic data lacks temporal information, this section focuses solely on spatial representation.
% %
% Calculating \( \boldsymbol{F}^I \) directly would require access to all local data, violating privacy constraints. To address this, we approximate the instrument-consistent representation using a moving average of the global model's feature \( \boldsymbol{F}_{g} \), inspired by the moving average technique for statistical approximation \cite{zhang2015deep}. Over multiple communication rounds, \( \boldsymbol{F}_{g} \) increasingly emphasizes instrument-consistent information as domain-specific variations, such as background diversity, gradually cancel out.
% %
% The approximated ground truth of \( \boldsymbol{F}^I \) derived from the moving average is then used to guide the pre-trained model through iterative updates, constraining its output to include only the instrument-consistent representation and achieving explicit quantification.
Directly incorporating synthetic data for training can harm performance due to the large domain gap between synthetic and real surgical data. Thus, we propose to explicitly quantify the instrument representation of synthetic data. Since synthetic data lacks temporal information, this section focuses solely on the spatial representation of instruments and ignores the RSC module. 
For simplicity, we denote \( \boldsymbol{F}_{ts} \) as \( \boldsymbol{F} \). 
% The pre-trained model \( v_{pre}^{0} \) is initially trained on the server using the synthetic dataset and then distributed to the sites as initialization. During the FL process, we will iteratively update this pre-trained model denoted as \( v_{pre}^{t} \).
%\cite{zhang2024eliminating, zhang2015deep} \cite{zhang2024eliminating}
As the domain representation \(\boldsymbol{D}\) and instrument consistent representation \(\boldsymbol{I}\) are distinct yet complementary components of the visual information in surgical videos, the former typically capture background and stylistic information, while the latter focuses on the surgical instruments. We define the representation \( \boldsymbol{F} \) extracted from the encoder as the addition of these two components: \( \boldsymbol{F} = \boldsymbol{D} + \boldsymbol{I} \). 

Furthermore, at each communication round $t$, we update the pre-trained model \( v^{t}_{pre} \) guided by the moving average of the global model's features \( \boldsymbol{F}_g^{t} \).
% We define the representation \( \boldsymbol{F} \) extracted from the encoder as the combination of domain representation and instrument-consistent representation: \( \boldsymbol{F} = \boldsymbol{D} + \boldsymbol{I} \) and constrain the iteratively updated pre-trained model to output only the instrument-consistent representation \( \boldsymbol I \) via a moving average of the global model's features \( \boldsymbol{F}_g \). 
% This approach works because, over multiple communication rounds, \( \boldsymbol{F}_g \) progressively emphasizes instrument-consistent representation as domain-specific variations, such as background diversity, gradually cancel out through iterative aggregation, all without requiring access to local data. Since synthetic data lacks temporal information, this section focuses solely on spatial representation.
%
This approach works because, over multiple communication rounds, the moving average mechanism ensures that the global feature \( \boldsymbol{F}_g^{t} \) progressively emphasizes the consistent instrument representation while reducing the influence of domain-specific variations, such as background diversity. During each round of aggregation, the consistent instrument representation, which remains stable and shared across sites, contributes cumulatively to \( \boldsymbol{F}_g \). In contrast, domain-specific variations, which are unique and often inconsistent across sites, tend to cancel out as they are averaged over multiple sites and communication rounds. 
% This iterative process enhances the focus on shared, consistent components—such as the instrument representation—without requiring direct access to local data. 
%Since synthetic data lacks temporal information, this section focuses solely on the spatial representation of instruments.

% Inspired by the widely used moving average technique for approximating statistics over batches of data \cite{zhang2015deep}, we can approximate the ground truth of instrument consistent representation by moving average of the feature \( \boldsymbol{F}_{g} \) from the global model. 

% To explicitly quantify the instrument consistent representation from the pre-trained model on synthesis data, 
% we define the representation \( \boldsymbol{f} \) extracted from the encoder to be the combination of bias representation and the instrument consistent representation representation: \( \boldsymbol{f}=\boldsymbol{f}^b+\boldsymbol{f}^g \). The biased representation includes background and stylistic information, whereas the instrument consistent representation is as same as the local side defined representation of instruments. 
Formally, the feature representation generated from the synthetic pre-trained model \( v^{t}_{pre} \) is expressed as \( \boldsymbol{F}^{t}_{sy} = \boldsymbol{D}^{t}_{sy} + \boldsymbol{I}^{t}_{sy} \). Similarly, we denote the feature representation generated by the global model  \( v^{t}_g \) encoder as \( \boldsymbol{F}^{t}_g \). Notably, \( \boldsymbol{I}^{t}_{sy} \) is a learnable domain descriptor, initialized randomly and continuously updated during the FL process.
% Formally, we begin by defining the feature representation \( \boldsymbol{F}^{t}_{sy} \) and \( {\boldsymbol{F}}^{t}_g \), which are generated by the pre-trained model \( v^{t}_{pre} \) encoder and the global model \( v^{t}_g \) encoder, respectively. 
% Thus, we have \( \boldsymbol{F}_{sy} = \boldsymbol{D}_{sy} + \boldsymbol{I}_{sy} \) 
% We consider the feature \( \boldsymbol{F}_{sy} \) from the synthetic pre-trained model to be composed of two components: the domain representation \( \boldsymbol{D}_{sy} \) and the instrument-consistent representation \( \boldsymbol{I}_{sy} \), such that \( \boldsymbol{F}_{sy} = \boldsymbol{D}_{sy} + \boldsymbol{I}_{sy} \). \(\boldsymbol{I}_{sy}\) is a learnable domain descriptor that is initialized randomly and is continuously updated across communication rounds during the FL process.
%
We use a moving average to iteratively update the \( \hat{\boldsymbol{F}_g} \), refining it across training rounds. The update rule is given by:
\begin{equation}
\label{fg}
\bm{\hat{F}}_g^{t+1} = (1 - \mu) \cdot \bm{\hat{F}}_g^{t} + \mu \cdot \bm{F}_g^{t+1},
\end{equation}

where \( \mu \) is the moving average coefficient; \( {\boldsymbol{F}_{g}^{t+1}} \) is the feature generated by the global model at communication round $t+1$. 
Explicitly quantify the instrument-consistent representation \( \boldsymbol{I}_{sy} \) or \( \boldsymbol{D}_{sy} \) are exactly the same, thus we fix the global model and update pre-trained model $v^{t}_{pre}$ by using the following objective:
\begin{equation}
\label{mse}
\mathcal{L}_{\text{pre}} = \mathcal{L}_{\text{Seg}} + \lambda_2 \cdot \frac{1}{N_e} \left\| \bm{F}^{t}_{\text{sy}} - \bm{D}^{t}_{\text{sy}} - \bm{\hat{F}}^{t+1}_g \right\|_2^2,
\end{equation}
where \( \lambda_2 \) is a regularization parameter; \( N_e \) is the number of elements in the feature  representation.
%接着这里可以解释下在simulated data做training时，instrument feature占主导；instrument feature的domain gap比较小，所以可以benefit
% The Cross-Entropy loss $\mathcal{L}_{\text{CE}}$ is given by:
% \begin{equation}
% \label{ceee}
% \mathcal{L}_{\text{CE}} = - \sum_{i=1}^N \mathcal{Y}_k(i) \log(P_k(i)) ,
% \end{equation}
% where \(P_k(i)\) represents the predicted probability for class \(i\) from the model with parameters \(\theta_k\), $\mathcal{Y}_k(i)$ is the ground truth class. 

\subsubsection{Synchronizing Local Convergence} 
% With the biased feature \( \boldsymbol{F}_{sy}^D \) explicitly quantified, we can synchronize the convergence in FL by removing this component, allowing the global model to learn the consistent instrument representation from the synthetic dataset. We define the knowledge transfer objective as follows:
With \( \boldsymbol{D}^{t+1}_{sy} \) explicitly quantified and pre-trained model updated to $v^{t+1}_{pre}$, we leverage the remaining \( \boldsymbol{I}^{t+1}_{sy} \), which contains the abundant instrument-consistent representation from the synthetic dataset, to enhance the global model and synchronize the convergence of different site models. By focusing solely on \( \boldsymbol{I}^{t+1}_{sy} \), the global model is empowered to learn a more robust and generalized instrument representation. This approach ensures that the global model achieves better generalization across diverse surgical environments and improves performance even on unseen sites. We fix the pre-trained model $v^{t+1}_{pre}$ and updated global model \( v^{t}_g \) as follows:
\begin{equation}
\label{kt}
\mathcal{L}_{g} = \mathcal{L}_{\mathrm{Seg}} + \lambda_3 \frac{1}{N_e} \left\| \boldsymbol{F}^{t}_g - (\boldsymbol{F}^{t+1}_{sy} - \boldsymbol{D}^{t+1}_{sy}) \right\|_2^2,
\end{equation}
where \( \lambda_3 \) is a regularization parameter, and \( N_e \) is the number of elements in the feature representation. The global model \( v^{t+1}_g \) is then distributed to local sites for further training.

\section{Experiments}

\subsection{Datasets and Training Implementation}

\subsubsection{Datasets}
\yj{We construct a new FL benchmark for surgical instrument segmentation to evaluate our method and promote FL research in surgical data science. Details and statistics of each site are shown in Table~\ref{FL_DATASET}.}
\yj{Each site in our FL setup corresponds to one full dataset. Specifically, Site A is EndoVis2017 \cite{endovis2017}, Site B is EndoVis2018 \cite{endovis2018}, and Site C is RARP \cite{rarp}, all from public surgical benchmarks. Site D (Hyst-YT) and Site E (Lob-YT) are additional datasets collected and annotated by us from open-source surgical videos on YouTube, following consistent part-level segmentation protocols.}
\yj{For Sites A to D, the training and testing splits follow the original protocol of each dataset, as shown in Table~\ref{FL_DATASET}. Site E is not involved in the federated training process and serves solely as an out-of-federation test set to assess model generalization.}
\yj{To further enrich the training corpus, we also employ the synthetic surgical dataset Sisvse \cite{siseve}, which is built in Unity using semantic image synthesis. This dataset offers diverse and privacy-free surgical scenes and is only used on server-side.}

\begin{table}[tp]
\centering
\scriptsize
\renewcommand{\arraystretch}{1}
\setlength\tabcolsep{2pt}
\caption{Dataset details and statistics for experimental evaluation.}
\resizebox{1\columnwidth}{!}{
\begin{tabular}{|c|c|c|}
\hline
\textbf{Dataset} & \textbf{Frames (Training / Testing)} & \textbf{Surgery Type} \\
\hline
A - EndoVis2017 \cite{endovis2017} & 1800 / 600 & Nephrectomy \\
\hline
B - EndoVis2018 \cite{endovis2018} & 2235 / 997 & Nephrectomy \\
\hline
C - RARP \cite{rarp} & 2023 / 1138 & Prostatectomy \\
\hline
D - Hyst-YT \cite{site4_1} & 998 / 982 & Hysterectomy \\
\hline
E - Lob-YT \cite{site4_1} & N/A / 203 & Lobectomy \\
\hline
Sythesis - Sisvse \cite{siseve} & 3600  / N/A & Synthetic Data \\
\hline
\end{tabular}
\label{FL_DATASET}
}
%\vspace{-3mm}
\end{table}

\begin{table*}[]
\scriptsize
\large
\centering
\caption{\yj{Comparison of state-of-the-art methods in terms of multiple metrics for sites A, B, C, D, and Avg. $\uparrow$indicates
the higher the score the better, and vice versa.}}
\label{comparison_combined}
\resizebox{\textwidth}{!}{%
\begin{tabular}{l||ccccc||ccccc}
\hline\hline
\multirow{2}{*}{Method} & \multicolumn{5}{c||}{\textbf{Dice}(\%)$\uparrow$} & \multicolumn{5}{c}{\textbf{IOU}(\%)$\uparrow$} \\ 
                        & A     & B     & C     & D     & Avg.  & A     & B     & C     & D     & Avg. \\ 
\hline
Local Train & 76.86 {\fontsize{10pt}{10pt}\selectfont$\pm$~0.35} & 69.92 {\fontsize{10pt}{10pt}\selectfont$\pm$~1.91} & 79.51 {\fontsize{10pt}{10pt}\selectfont$\pm$~0.70} & 59.72 {\fontsize{10pt}{10pt}\selectfont$\pm$~1.39} & 71.50 {\fontsize{10pt}{10pt}\selectfont$\pm$~0.91} & 63.55 {\fontsize{10pt}{10pt}\selectfont$\pm$~0.58} & 54.11 {\fontsize{10pt}{10pt}\selectfont$\pm$~0.75} & 66.82 {\fontsize{10pt}{10pt}\selectfont$\pm$~0.43} & 44.67 {\fontsize{10pt}{10pt}\selectfont$\pm$~1.23} & 57.29 {\fontsize{10pt}{10pt}\selectfont$\pm$~0.46} \\

Upperbound & 85.99 {\fontsize{10pt}{10pt}\selectfont$\pm$~1.90} & 83.72 {\fontsize{10pt}{10pt}\selectfont$\pm$~0.68} & 89.63 {\fontsize{10pt}{10pt}\selectfont$\pm$~0.71} & 73.50 {\fontsize{10pt}{10pt}\selectfont$\pm$~1.54} & 83.21 {\fontsize{10pt}{10pt}\selectfont$\pm$~0.85} & 75.94 {\fontsize{10pt}{10pt}\selectfont$\pm$~3.17} & 72.97 {\fontsize{10pt}{10pt}\selectfont$\pm$~0.82} & 81.53 {\fontsize{10pt}{10pt}\selectfont$\pm$~1.14} & 58.95 {\fontsize{10pt}{10pt}\selectfont$\pm$~2.33} & 72.35 {\fontsize{10pt}{10pt}\selectfont$\pm$~1.03} \\
\hline
FedLD \cite{zeng2024tackling} & 80.19 {\fontsize{10pt}{10pt}\selectfont$\pm$~0.11} & 78.68 {\fontsize{10pt}{10pt}\selectfont$\pm$~0.51} & 80.58 {\fontsize{10pt}{10pt}\selectfont$\pm$~0.71} & 70.69 {\fontsize{10pt}{10pt}\selectfont$\pm$~0.53} & 77.53 {\fontsize{10pt}{10pt}\selectfont$\pm$~0.24} & 67.96 {\fontsize{10pt}{10pt}\selectfont$\pm$~0.07} & 65.89 {\fontsize{10pt}{10pt}\selectfont$\pm$~0.40} & 67.78 {\fontsize{10pt}{10pt}\selectfont$\pm$~0.27} & 56.17 {\fontsize{10pt}{10pt}\selectfont$\pm$~0.45} & 64.45 {\fontsize{10pt}{10pt}\selectfont$\pm$~0.26} \\

FedDP \cite{feddp} & 
78.91 {\fontsize{10pt}{10pt}\selectfont$\pm$~0.32} & 
81.24 {\fontsize{10pt}{10pt}\selectfont$\pm$~0.42} & 
85.62 {\fontsize{10pt}{10pt}\selectfont$\pm$~0.18} & 
71.58 {\fontsize{10pt}{10pt}\selectfont$\pm$~0.25} & 
79.34 {\fontsize{10pt}{10pt}\selectfont$\pm$~0.44} & 
66.45 {\fontsize{10pt}{10pt}\selectfont$\pm$~0.39} & 
68.89 {\fontsize{10pt}{10pt}\selectfont$\pm$~0.38} & 
74.74 {\fontsize{10pt}{10pt}\selectfont$\pm$~0.21} & 
53.32 {\fontsize{10pt}{10pt}\selectfont$\pm$~0.29} & 
65.85 {\fontsize{10pt}{10pt}\selectfont$\pm$~0.45} \\

FedAvg \cite{fedavg} & 80.81 {\fontsize{10pt}{10pt}\selectfont$\pm$~0.40} & 81.44 {\fontsize{10pt}{10pt}\selectfont$\pm$~0.45} & 83.70 {\fontsize{10pt}{10pt}\selectfont$\pm$~0.01} & 74.01 {\fontsize{10pt}{10pt}\selectfont$\pm$~0.83} & 79.99 {\fontsize{10pt}{10pt}\selectfont$\pm$~0.62} & 68.47 {\fontsize{10pt}{10pt}\selectfont$\pm$~0.37} & 69.49 {\fontsize{10pt}{10pt}\selectfont$\pm$~0.43} & 72.93 {\fontsize{10pt}{10pt}\selectfont$\pm$~0.46} & 59.56 {\fontsize{10pt}{10pt}\selectfont$\pm$~0.33} & 67.61 {\fontsize{10pt}{10pt}\selectfont$\pm$~0.50} \\

FedRep \cite{fedrep} & 81.15 {\fontsize{10pt}{10pt}\selectfont$\pm$~2.04} & 82.16 {\fontsize{10pt}{10pt}\selectfont$\pm$~0.74} & 87.11 {\fontsize{10pt}{10pt}\selectfont$\pm$~1.12} & 70.20 {\fontsize{10pt}{10pt}\selectfont$\pm$~2.10} & 80.66 {\fontsize{10pt}{10pt}\selectfont$\pm$~0.66} & 67.01 {\fontsize{10pt}{10pt}\selectfont$\pm$~4.39} & 70.76 {\fontsize{10pt}{10pt}\selectfont$\pm$~1.21} & 77.33 {\fontsize{10pt}{10pt}\selectfont$\pm$~2.15} & 55.61 {\fontsize{10pt}{10pt}\selectfont$\pm$~2.55} & 67.68 {\fontsize{10pt}{10pt}\selectfont$\pm$~0.91} \\
FedAS \cite{yang2024fedas} & 84.28 {\fontsize{10pt}{10pt}\selectfont$\pm$~0.46} & 82.73 {\fontsize{10pt}{10pt}\selectfont$\pm$~0.40} & 88.61 {\fontsize{10pt}{10pt}\selectfont$\pm$~0.18} & 70.15 {\fontsize{10pt}{10pt}\selectfont$\pm$~2.12} & 81.44 {\fontsize{10pt}{10pt}\selectfont$\pm$~0.50} & 73.96 {\fontsize{10pt}{10pt}\selectfont$\pm$~0.11} & 71.97 {\fontsize{10pt}{10pt}\selectfont$\pm$~0.06} & 80.07 {\fontsize{10pt}{10pt}\selectfont$\pm$~0.03} & 54.80 {\fontsize{10pt}{10pt}\selectfont$\pm$~2.65} & 70.20 {\fontsize{10pt}{10pt}\selectfont$\pm$~0.56} \\

FedST (Ours)           & \textbf{84.73} {\fontsize{10pt}{10pt}\selectfont$\pm$~0.87} & \textbf{83.15} {\fontsize{10pt}{10pt}\selectfont$\pm$~0.25} & \textbf{89.13} {\fontsize{10pt}{10pt}\selectfont$\pm$~0.40} & \textbf{71.46} {\fontsize{10pt}{10pt}\selectfont$\pm$~3.29} & \textbf{82.62} {\fontsize{10pt}{10pt}\selectfont$\pm$~0.09} & \textbf{74.38} {\fontsize{10pt}{10pt}\selectfont$\pm$~1.23} & \textbf{72.07} {\fontsize{10pt}{10pt}\selectfont$\pm$~0.34} & \textbf{80.71} {\fontsize{10pt}{10pt}\selectfont$\pm$~0.58} & \textbf{56.38} {\fontsize{10pt}{10pt}\selectfont$\pm$~4.00} & \textbf{71.47} {\fontsize{10pt}{10pt}\selectfont$\pm$~0.07} \\
\hline\hline
\multirow{2}{*}{Method} & \multicolumn{5}{c||}{\textbf{HD95}(\textit{pix.})$\downarrow$} & \multicolumn{5}{c}{\textbf{ASSD}(\textit{pix.})$\downarrow$} \\ 
                        & A     & B     & C     & D     & Avg.  & A     & B     & C     & D     & Avg. \\ 
\hline
% 以下内容保留原始格式，等待你补充 std
Local Train & 186.21 {\fontsize{10pt}{10pt}\selectfont$\pm$~3.57} & 291.52 {\fontsize{10pt}{10pt}\selectfont$\pm$~1.98} & 258.65 {\fontsize{10pt}{10pt}\selectfont$\pm$~2.01} & 596.08 {\fontsize{10pt}{10pt}\selectfont$\pm$~6.34} & 333.12 {\fontsize{10pt}{10pt}\selectfont$\pm$~1.69} & 49.01 {\fontsize{10pt}{10pt}\selectfont$\pm$~1.95} & 88.63 {\fontsize{10pt}{10pt}\selectfont$\pm$~1.97} & 64.57 {\fontsize{10pt}{10pt}\selectfont$\pm$~0.93} & 168.41 {\fontsize{10pt}{10pt}\selectfont$\pm$~4.04} & 92.66 {\fontsize{10pt}{10pt}\selectfont$\pm$~1.25} \\

Upperbound & 94.72 {\fontsize{10pt}{10pt}\selectfont$\pm$~24.71} & 119.04 {\fontsize{10pt}{10pt}\selectfont$\pm$~14.63} & 116.01 {\fontsize{10pt}{10pt}\selectfont$\pm$~1.16} & 434.84 {\fontsize{10pt}{10pt}\selectfont$\pm$~39.18} & 191.65 {\fontsize{10pt}{10pt}\selectfont$\pm$~12.75} & 19.71 {\fontsize{10pt}{10pt}\selectfont$\pm$~3.25} & 32.78 {\fontsize{10pt}{10pt}\selectfont$\pm$~4.30} & 29.04 {\fontsize{10pt}{10pt}\selectfont$\pm$~0.66} & 137.54 {\fontsize{10pt}{10pt}\selectfont$\pm$~13.27} & 54.77 {\fontsize{10pt}{10pt}\selectfont$\pm$~2.92} \\

\hline
FedLD \cite{zeng2024tackling} & 
148.50 {\fontsize{10pt}{10pt}\selectfont$\pm$~1.36} & 
182.85 {\fontsize{10pt}{10pt}\selectfont$\pm$~0.91} & 
248.26 {\fontsize{10pt}{10pt}\selectfont$\pm$~2.27} & 
550.65 {\fontsize{10pt}{10pt}\selectfont$\pm$~1.50} & 
282.56 {\fontsize{10pt}{10pt}\selectfont$\pm$~1.30} & 
41.09 {\fontsize{10pt}{10pt}\selectfont$\pm$~1.06} & 
51.03 {\fontsize{10pt}{10pt}\selectfont$\pm$~1.11} & 
60.58 {\fontsize{10pt}{10pt}\selectfont$\pm$~0.55} & 
175.27 {\fontsize{10pt}{10pt}\selectfont$\pm$~1.77} & 
81.99 {\fontsize{10pt}{10pt}\selectfont$\pm$~0.80} \\

FedDP \cite{feddp} & 
137.23 {\fontsize{10pt}{10pt}\selectfont$\pm$~1.40} & 
130.16 {\fontsize{10pt}{10pt}\selectfont$\pm$~1.01} & 
149.96 {\fontsize{10pt}{10pt}\selectfont$\pm$~0.51} & 
477.46 {\fontsize{10pt}{10pt}\selectfont$\pm$~1.08} & 
223.70 {\fontsize{10pt}{10pt}\selectfont$\pm$~15.26} & 
26.13 {\fontsize{10pt}{10pt}\selectfont$\pm$~0.63} & 
35.28 {\fontsize{10pt}{10pt}\selectfont$\pm$~0.64} & 
36.91 {\fontsize{10pt}{10pt}\selectfont$\pm$~0.64} & 
151.18 {\fontsize{10pt}{10pt}\selectfont$\pm$~1.07} & 
62.37 {\fontsize{10pt}{10pt}\selectfont$\pm$~1.61} \\

FedAvg \cite{fedavg} & 140.44 {\fontsize{10pt}{10pt}\selectfont$\pm$~0.55} & 179.42 {\fontsize{10pt}{10pt}\selectfont$\pm$~0.37} & 216.03 {\fontsize{10pt}{10pt}\selectfont$\pm$~0.66} & 461.03 {\fontsize{10pt}{10pt}\selectfont$\pm$~0.62} & 249.73 {\fontsize{10pt}{10pt}\selectfont$\pm$~1.82} & 29.51 {\fontsize{10pt}{10pt}\selectfont$\pm$~0.56} & 54.22 {\fontsize{10pt}{10pt}\selectfont$\pm$~1.06} & 52.20 {\fontsize{10pt}{10pt}\selectfont$\pm$~0.46} & 134.19 {\fontsize{10pt}{10pt}\selectfont$\pm$~0.44} & 67.53 {\fontsize{10pt}{10pt}\selectfont$\pm$~2.18} \\

FedRep \cite{fedrep} & 
128.50 {\fontsize{10pt}{10pt}\selectfont$\pm$~2.04} & 
125.30 {\fontsize{10pt}{10pt}\selectfont$\pm$~1.81} & 
158.93 {\fontsize{10pt}{10pt}\selectfont$\pm$~1.33} & 
465.18 {\fontsize{10pt}{10pt}\selectfont$\pm$~0.70} & 
219.48 {\fontsize{10pt}{10pt}\selectfont$\pm$~1.02} & 
27.30 {\fontsize{10pt}{10pt}\selectfont$\pm$~0.32} & 
36.51 {\fontsize{10pt}{10pt}\selectfont$\pm$~1.46} & 
38.61 {\fontsize{10pt}{10pt}\selectfont$\pm$~1.52} & 
142.29 {\fontsize{10pt}{10pt}\selectfont$\pm$~0.13} & 
61.18 {\fontsize{10pt}{10pt}\selectfont$\pm$~1.90} \\

FedAS \cite{yang2024fedas} & 112.27 {\fontsize{10pt}{10pt}\selectfont$\pm$~1.60} & 125.13 {\fontsize{10pt}{10pt}\selectfont$\pm$~0.84} & 134.57 {\fontsize{10pt}{10pt}\selectfont$\pm$~0.57} & 399.65 {\fontsize{10pt}{10pt}\selectfont$\pm$~2.03} & 192.41 {\fontsize{10pt}{10pt}\selectfont$\pm$~1.38} & 24.13 {\fontsize{10pt}{10pt}\selectfont$\pm$~0.29} & 39.50 {\fontsize{10pt}{10pt}\selectfont$\pm$~1.27} & 33.09 {\fontsize{10pt}{10pt}\selectfont$\pm$~1.01} & 126.62 {\fontsize{10pt}{10pt}\selectfont$\pm$~1.95} & 55.84 {\fontsize{10pt}{10pt}\selectfont$\pm$~2.01} \\

FedST (Ours) & \textbf{90.67} {\fontsize{10pt}{10pt}\selectfont$\pm$~0.49} & \textbf{124.90} {\fontsize{10pt}{10pt}\selectfont$\pm$~0.70} & \textbf{123.27} {\fontsize{10pt}{10pt}\selectfont$\pm$~2.35} & \textbf{404.12} {\fontsize{10pt}{10pt}\selectfont$\pm$~4.12} & \textbf{185.74} {\fontsize{10pt}{10pt}\selectfont$\pm$~1.45} & \textbf{19.20} {\fontsize{10pt}{10pt}\selectfont$\pm$~0.15} & \textbf{34.28} {\fontsize{10pt}{10pt}\selectfont$\pm$~0.09} & \textbf{28.97} {\fontsize{10pt}{10pt}\selectfont$\pm$~0.42} & \textbf{122.20} {\fontsize{10pt}{10pt}\selectfont$\pm$~0.83} & \textbf{51.16} {\fontsize{10pt}{10pt}\selectfont$\pm$~0.24} \\
\hline\hline

\end{tabular}%
}
\end{table*}

\begin{table}[htbp]
\centering
\large
\renewcommand{\arraystretch}{1}
\setlength\tabcolsep{10pt}
\caption{\yj{Detailed performance metrics for Site E (Out Fed)}}
\resizebox{1\columnwidth}{!}{
\begin{tabular}{l|cccc}
\hline\hline
Method & \textbf{Dice(\%)$\uparrow$}  & \textbf{IOU(\%)$\uparrow$}  & \textbf{HD95(\textit{pix.})$\downarrow$} & \textbf{ASSD(\textit{pix.})$\downarrow$}  \\ 
\hline
Local Train & 26.99 {\fontsize{10pt}{10pt}\selectfont$\pm$~0.03} & 16.97 {\fontsize{10pt}{10pt}\selectfont$\pm$~0.55} & 593.59 {\fontsize{10pt}{10pt}\selectfont$\pm$~3.06} & 231.97 {\fontsize{10pt}{10pt}\selectfont$\pm$~4.46} \\
Upperbound & 27.56 {\fontsize{10pt}{10pt}\selectfont$\pm$~0.48} & 17.05 {\fontsize{10pt}{10pt}\selectfont$\pm$~0.05} & 517.38 {\fontsize{10pt}{10pt}\selectfont$\pm$~1.34} & 180.32 {\fontsize{10pt}{10pt}\selectfont$\pm$~0.69} \\
\hline
FedDP \cite{feddp} & 26.60 {\fontsize{10pt}{10pt}\selectfont$\pm$~1.73} & 16.69 {\fontsize{10pt}{10pt}\selectfont$\pm$~1.31} & 652.57 {\fontsize{10pt}{10pt}\selectfont$\pm$~39.67} & 251.26 {\fontsize{10pt}{10pt}\selectfont$\pm$~20.29} \\
FedAS \cite{yang2024fedas} & 27.87 {\fontsize{10pt}{10pt}\selectfont$\pm$~0.22} & 16.95 {\fontsize{10pt}{10pt}\selectfont$\pm$~0.17} & 495.55 {\fontsize{10pt}{10pt}\selectfont$\pm$~7.99} & 183.73 {\fontsize{10pt}{10pt}\selectfont$\pm$~8.91} \\
FedLD \cite{zeng2024tackling} & 
28.74 {\fontsize{10pt}{10pt}\selectfont$\pm$~0.43} & 
17.84 {\fontsize{10pt}{10pt}\selectfont$\pm$~0.20} & 
509.95 {\fontsize{10pt}{10pt}\selectfont$\pm$~1.10} & 
201.19 {\fontsize{10pt}{10pt}\selectfont$\pm$~0.56} \\
FedAvg \cite{fedavg} & 28.73 {\fontsize{10pt}{10pt}\selectfont$\pm$~0.56} & 17.76 {\fontsize{10pt}{10pt}\selectfont$\pm$~0.23} & 635.64 {\fontsize{10pt}{10pt}\selectfont$\pm$~0.73} & 223.42 {\fontsize{10pt}{10pt}\selectfont$\pm$~1.39} \\
FedRep \cite{fedrep} & 28.76 {\fontsize{10pt}{10pt}\selectfont$\pm$~0.40} & 17.75 {\fontsize{10pt}{10pt}\selectfont$\pm$~0.26} & 453.75 {\fontsize{10pt}{10pt}\selectfont$\pm$~8.24} & 165.61 {\fontsize{10pt}{10pt}\selectfont$\pm$~1.49} \\
\textbf{FedST (Ours)} & \textbf{77.97 {\fontsize{10pt}{10pt}\selectfont$\pm$~0.68}} & \textbf{64.31 {\fontsize{10pt}{10pt}\selectfont$\pm$~0.79}} & \textbf{310.91 {\fontsize{10pt}{10pt}\selectfont$\pm$~2.37}} & \textbf{81.42 {\fontsize{10pt}{10pt}\selectfont$\pm$~2.28}} \\
\hline\hline
\end{tabular}
\label{tab:site_e_metrics}
}
\end{table}

% \begin{table}[tp]
% \centering
% \large
% \renewcommand{\arraystretch}{1}
% \setlength\tabcolsep{10pt}
% \caption{Detailed performance metrics for Site E (Out Fed)}
% \resizebox{1\columnwidth}{!}{
% \begin{tabular}{l|cccc}
% \hline\hline
% Method & \textbf{Dice(\%)$\uparrow$}  & \textbf{IOU(\%)$\uparrow$}  & \textbf{HD95(\textit{pix.})$\downarrow$} & \textbf{ASSD(\textit{pix.})$\downarrow$}  \\ 
% \hline
% Local Train            & 26.98 & 16.65 & 595.80 & 234.57 \\
% Centralized            & 27.90 & 17.08 & 516.41 & 179.81 \\ \hline
% FedDP     \cite{feddp}                  & 27.61 & 17.35 & 600.21 & 229.64 \\
% FedRep    \cite{fedrep}               & 28.04 & 17.54 & 544.59 & 208.49 \\
% FedAS     \cite{yang2024fedas}            & 28.04 & 17.09 & 484.46 & 173.01 \\  
% FedLD     \cite{zeng2024tackling}              & 28.43 & 17.69 & 510.75 & 201.60 \\
% FedAvg    \cite{fedavg}            & 28.80 & 18.00 & 634.63 & 211.60 \\ \hline
% FedST (Ours)          & \textbf{76.99} & \textbf{63.29} & \textbf{313.27} & \textbf{88.62} \\
% \hline\hline
% \end{tabular}
% \label{tab:site_e_metrics}
% }
% \end{table}

\begin{figure*}[t]
    \centering
    \includegraphics[width=\textwidth]{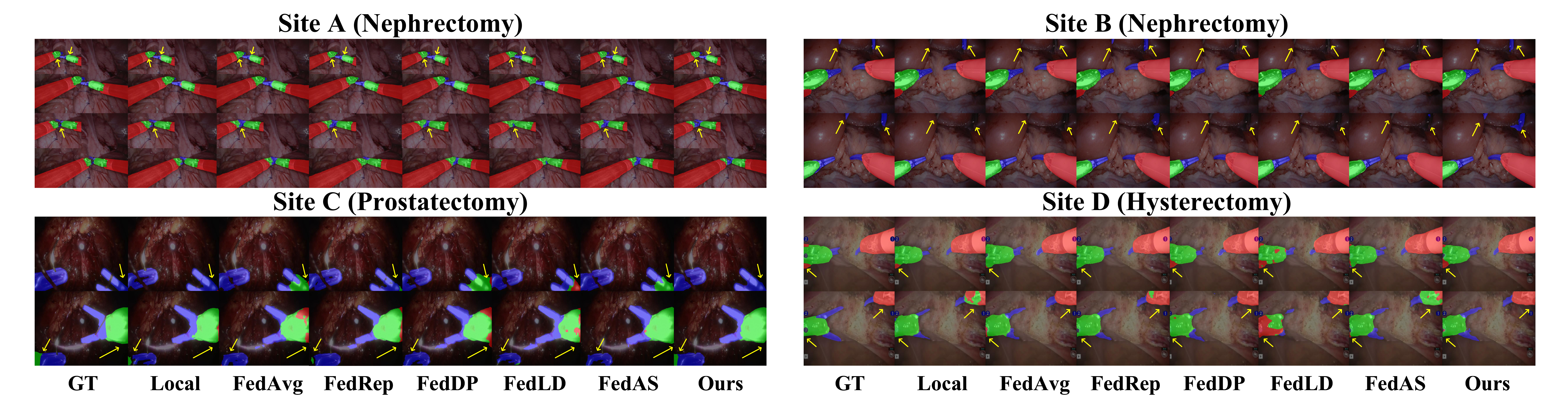}
    \caption{Qualitative comparisons of different methods on two continuous video frames of varying surgery types from site A to Site D. Ground truths, and part segmentation results (shaft, wrist, and jaws) from local train, FedAvg, FedRep, FedAS, and our FedST are illustrated.}
    \label{vis}
\end{figure*}

\begin{figure}[t]
    \centering
    \includegraphics[width=\columnwidth]{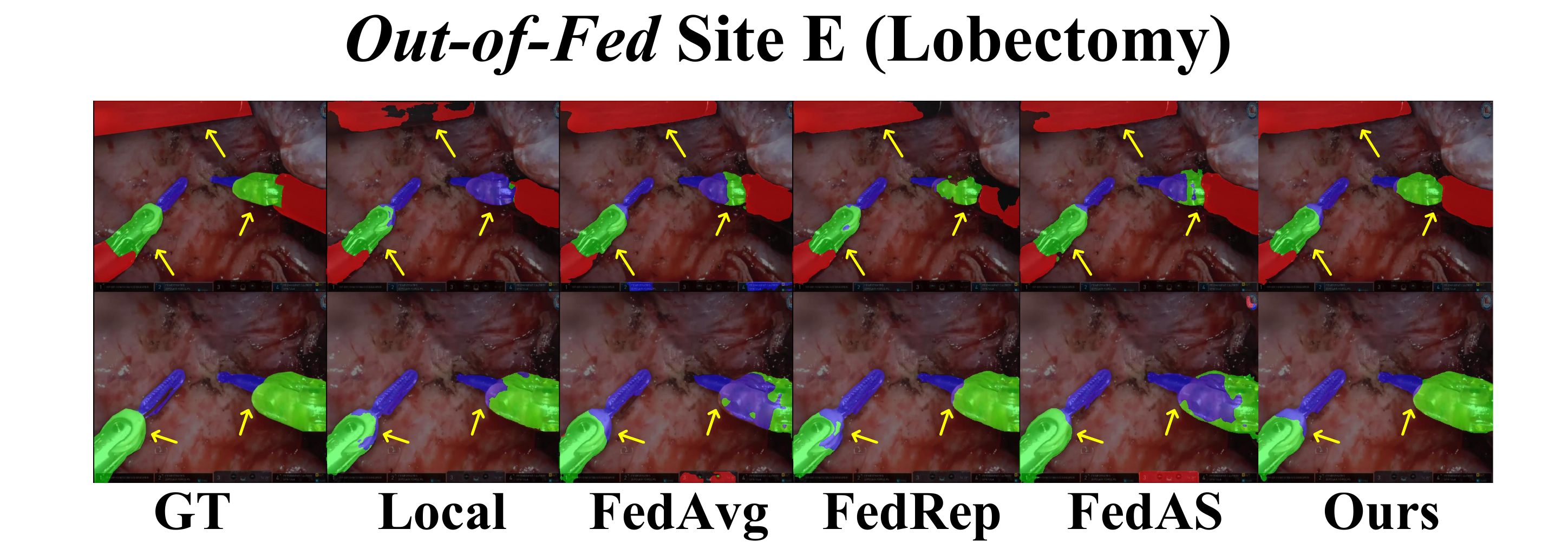}
    \vspace*{-0.6cm}
    \caption{Qualitative comparisons of different methods on two continuous video frames from site E. Ground truths and part segmentation results (inference only) from local train, FedAvg, FedRep, FedAS, and our FedST are illustrated.}
    \label{vis_5}
\end{figure}

% \begin{table}[t]
% \centering
% \scriptsize
% \renewcommand{\arraystretch}{1}
% \setlength\tabcolsep{2pt}
% \caption{Ablation Study of Key Method Components on Dice.}
% \resizebox{1\columnwidth}{!}{
% \begin{tabular}{ccc||cccccc}
% \hline
% \hline
% \multirow{2}{*}{TS} & \multirow{2}{*}{Prompt} & \multirow{2}{*}{SERQ} & \multicolumn{6}{c}{\textbf{Dice}(\%)$\uparrow$} \\ 
%      &       &          & A         & B     & C     & D     & Avg.  & E (Out Fed) \\ \hline
%              &          &           & 74.66 & 78.68 & 82.93 & 65.69 & 75.49 & 27.61 \\ 
% \ding{51}    &          &           & 83.96 & 80.72 & 86.43 & 71.86 & 80.74 & 71.19 \\ 
% \ding{51}    & \ding{51}&           & 83.23 & 81.09 & 87.87 & \textbf{75.41} & 81.90 & 72.21 \\ 
%              &          & \ding{51} & 84.25 & 78.98 & 85.41 & 69.01 & 79.41 & 28.23 \\ 
% \ding{51}    &          & \ding{51} & \textbf{84.31} & 81.52 & 88.27 & 74.06 & 82.04 & 76.78 \\ 
% \ding{51}    & \ding{51}   & \ding{51} & 83.56 & \textbf{82.97} & \textbf{89.38} & 74.85 & \textbf{82.69} & \textbf{76.99} \\  
% \hline
% \hline
% \end{tabular}
% \label{ablation1}
% }
% \end{table}

\begin{table}[t]
\centering
\scriptsize
\renewcommand{\arraystretch}{1}
\setlength\tabcolsep{2pt}
\caption{\yj{Ablation Study of Key Method Components on Dice.}}
\resizebox{1\columnwidth}{!}{
\begin{tabular}{cccc|cccccc}
\hline
\hline
\multirow{2}{*}{TS} & \multirow{2}{*}{Prompt} & \multirow{2}{*}{CS} & \multirow{2}{*}{SERQ} & \multicolumn{6}{c}{\textbf{Dice} (\%)$\uparrow$} \\
 & & & & A & B & C & D & Avg. & E (Out Fed) \\
\hline
\hline
 &      &  &  & 74.66 & 78.68 & 82.93 & 65.69 & 75.49 & 27.61 \\ 
\ding{51} & & & & 83.96 & 80.72 & 86.43 & 71.86 & 80.74 & 71.19 \\ 
\ding{51} & \ding{51} &   &      & 83.87 & 82.31 & 89.20 & 69.64 & 81.26 & 72.22 \\
\ding{51} & \ding{51} & \ding{51} &      & 83.23 & 81.09 & 87.87 & \textbf{75.41} & 81.90 & 72.21 \\ 
 &  &  & \ding{51} & 84.25 & 78.98 & 85.41 & 69.01 & 79.41 & 28.23 \\ 
\ding{51} &   &  & \ding{51}  & \textbf{84.31} & 81.52 & 88.27 & 74.06 & 82.04 & 76.78 \\ 
\ding{51} & \ding{51} & \ding{51} & \ding{51} & 83.56 & \textbf{82.97} & \textbf{89.38} & 74.85 & \textbf{82.69} & \textbf{76.99} \\
\hline
\hline
\end{tabular}
\label{ablation1}
}
\end{table}

\begin{table}[t]
\centering
\scriptsize
\renewcommand{\arraystretch}{1}
\setlength\tabcolsep{2pt}
\caption{Comparison of personalizing different layers in the self-attention networks.}
\label{fl_setting}
\resizebox{1\columnwidth}{!}{
\begin{tabular}{|c|c|c|cccccc|}
\hline
\multicolumn{1}{|c|}{Encoder} & \multicolumn{2}{c|}{Decoder} & \multicolumn{6}{c|}{\textbf{Dice (\%)}~($\uparrow$)} \\ \hline
\multicolumn{1}{|c|}{Attn.q}   & Attn.q          & Other       & A     & B     & C     & D     & Avg.   & E (Out Fed) \\ \hline
                               &                 &             & $\bm{84.69}$ & $\bm{83.15}$  & 87.54          & 73.40          & 82.19          & 76.42               \\ \hline
\multicolumn{1}{|c|}{$\Delta$} & $\Delta$        & $\Delta$    & 84.11        & 81.94         & 87.89          & 72.30          & 81.56          & 10.25               \\ \hline
                               & $\Delta$        &             & 83.59        & 80.71         & 87.79          & 73.33          & 81.36          & 56.23               \\ \hline
\multicolumn{1}{|c|}{$\Delta$} & $\Delta$        &             & 83.34        & 82.36         & 89.38          & 71.53          & 81.65          & 54.37               \\ \hline
\multicolumn{1}{|c|}{$\Delta$} &                 &             & 83.56        & 82.97         & $\bm{89.38}$   & $\bm{74.85}$   & $\bm{82.69}$   & $\bm{76.99}$         \\ \hline
\end{tabular}
}
\begin{flushleft}
\footnotesize Note: The symbol $\Delta$ indicates parameters that are personalized during FL.
\end{flushleft}
\end{table}

\begin{figure*}[t]
    \centering
    % \hspace*{-1cm}
    \includegraphics[width=\textwidth]{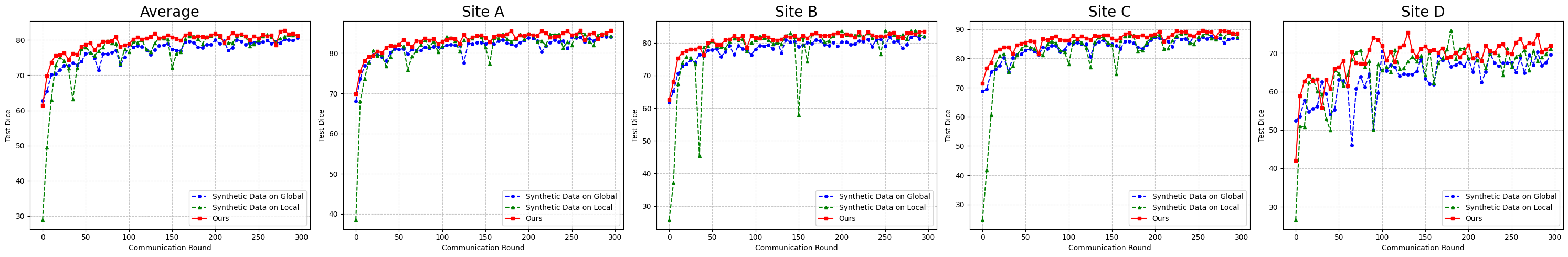}
    \caption{Learning curves of different methods utilizing synthetic data, represented by different colors. Test Dice scores versus the communication rounds for the average and each participated site are shown from left to right.}
    \label{vis_6}
    
\end{figure*}

\subsubsection{Metrics}
We employ several metrics to evaluate segmentation performance: the Dice similarity coefficient, Intersection over Union (IOU), Average Symmetric Surface Distance (ASSD), and Hausdorff distance-95 (HD95). These metrics collectively provide a comprehensive assessment of segmentation quality, addressing both shape conformity and spatial accuracy \cite{feddp}. 

\subsubsection{Implementation Details}
Our FedST employs a Pyramid Vision Transformer (PVT) \cite{wang2022pvt} as the backbone of the encoder. 
For the local site training, the network is optimized with a learning rate of $6\times10^{-5}$ by the AdamW optimizer. 
Each video clip consists of the current frame and several precedent frames sampled at an interval. \yj{To ensure computational efficiency, we empirically limit the input to four frames—comprising the current frame and its three immediate historical frames—with a temporal interval of 0.2 seconds. For the initial frames where sufficient history is unavailable, we duplicate the current frame to fill the missing historical slots.}

The receptive field, pooling kernel, and window size are empirically set as follows: \( \boldsymbol{r} = \{49, 20, 6, 7\} \), \( \boldsymbol{p} = \{7, 4, 2, 1\} \), and \( s = 7 \).
The communication with global server is performed after every 100 training iterations in local sites until communication round $T$ achieves 300.
For the model update on the global server during communication, we use 100 sample batches with a learning rate of $1\times10^{-4}$ for each round of model update on the server side. 
The network is optimized using Adam optimizer with a batch size of 2. 
We set \( \lambda_1 = \lambda_2 = \lambda_3 = 0.3 \) and \( \mu = 0.5 \).
In each site or server training, data augmentation (rotation and scale) is performed. 
\yj{For a fair comparison, all state-of-the-art methods were trained with the same data augmentation strategies, pre-trained on the same synthetic datasets, and optimized using identical training schedules—including the same optimizer settings and 300,000 training iterations.}
\yj{Unless otherwise specified, all experiments without the SERQ module do not utilize any synthetic data during training.
}
All experiments are performed on a single NVIDIA RTX A5000 GPU. \yj{To ensure statistical reliability, each experiment is repeated three times with different random seeds, and the reported results are presented as the mean and standard deviation.}

\subsection{Comparison With State-of-the-Art Methods}
\subsubsection{Internal Validation}
To validate the superiority of our FedST, we first conduct the internal validation (site A to D) with the general FL methods (FedAvg \cite{fedavg}), and several state-of-the-art PFL methods (FedRep \cite{fedrep}, FedDP \cite{feddp}, FedLD \cite{zeng2024tackling}, FedAS \cite{yang2024fedas}). We also locally train the models by only using their own datasets in different sites without the FL technique as the baseline (Local Train). Meanwhile, the model is trained by centralizing data from all sites as the upperbond. 

The results are shown in Table \ref{comparison_combined}. It can be found that FL methods show better performance than local train (Dice 70.98\% vs. 75.49\%). This indicates that the FL is necessary for the surgical instrument segmentation task. Data centralization achieves the best performance, indicating that collaborative training from multiple sites is beneficial. However, this method overlooks privacy concerns, limiting its utilization in real-world deployments. Our FedST can achieve promising performance while keeping privacy protection, which consistently outperforms other state-of-the-art FL methods across all evaluation metrics in the average values. This is attributed to the sharing and enhancement of consistent instrument representation across sites via RSC and SERQ modules, which strengthens the global model. These contributions are particularly beneficial for sites with limited data, such as Site D, even outperforming the centralized approach. 

We visualize the segmentation results from different methods in Fig. \ref{vis}, further demonstrating the superiority of our FedST. We can see that the local training approach and other methods frequently succumb to misclassification errors, such as incorrectly categorizing instrument parts (e.g., classifying parts of the wrist as other components). In contrast, our FedST predicts more accurate segmentation across different sites, unaffected by the variations in background. Furthermore, our FedST maintains consistent results across consecutive frames benefiting from temporal modeling, while other methods struggle to preserve such consistency.

\subsubsection{External Validation}
To evaluate the robustness of different methods, we further conduct external validation on the out-of-federation site E. 
\yj{We aggregate the parameters of personalized models on the server via weighted averaging and use the resulting global model for inference, which requires only a single forward pass and avoids the computational overhead of model ensembling.}
As shown in Table \ref{tab:site_e_metrics}, other methods nearly fail to generalize, showing considerably lower performance. Instead, without any additional adaption on Site E, our FedST achieves promising performance. This underscores that by decoupling temporal information, our method can separate the consistent instrument representation from background tissue, and therefore largely enhance the generalization capability to the surgical scenario even with distinct background appearance. 

Regarding the visual comparison on this out-of-federation setting, we show the results on Site E shown in Fig.\ref{vis_5}. We find that other methods tend to misclassify. With enhanced instrument-consistent representation, our model can correctly segment instruments, which shows stronger generalization across diverse surgical scenarios.

\subsection{Ablation Studies}
\subsubsection{Effectiveness of Key Components}
%先要把setting有几个，分别怎么configuration的写出来
%ablation的顺序最好和method部分介绍的顺序一致
We conduct ablation experiments to validate the effectiveness of different key components in the proposed method and obtain six configurations. The results are presented in Table \ref{ablation1}: \textbf{1st row}: we train the pure FedDP model as the baseline, where no RSC and SERQ are involved. \textbf{2nd row}: the temporal modeling and separation step (TS) of RSC is utilized to separate the consistent representation of each site in FL. 
\yj{\textbf{3rd row}: we include the textual prompt embedding without the channel selection network (CS), aiming to provide lightweight site-specific guidance. \textbf{4th row}: we further include CS to form a more complete RSC module, enabling the model to selectively enhance informative channels for each site.}
% \textbf{3rd row}: we include the textual guidance to form the full RSC to enhance the representation of each site further. 
\yj{\textbf{5th row}: only SERQ which explicitly quantifies synthetic dataset is utilized to facilitate the convergence in FL model fusion without any temporal modeling and separation. 
\textbf{6th row}: adding TS with SERQ to train the model.}
\textbf{7th row}: our proposed full FedST. 

The results in Table \ref{ablation1} clearly show the importance of each component in FedST. In the baseline model (1st row), where neither RSC nor SERQ is used, the performance is suboptimal, particularly on site E (27.61\%). When the temporal modeling and decoupling step of RSC is introduced (2nd row), the performance improves, indicating that this strategy can help separate site-specific representations and address background diversity more effectively. \yj{Adding only the textual prompt (3rd row) does not lead to significant improvement due to the lack of explicit feature selection. In contrast, introducing the channel selection network (4th row) significantly enhances the performance, particularly on sites D, demonstrating its ability to guide each site in attending to its most relevant feature channels. This highlights that channel selection serves as an effective personalized mechanism to amplify the representation learning for each local site.} 
In the 5th row, introducing the SERQ module leverages the instrument representation from the synthetic dataset to enhance the global model, ensuring convergence during model fusion and stabilizing the optimization process, especially on site D with limited data. 
\yj{In the 6th row of Table~\ref{ablation1}, incorporating the temporal modeling leads to consistent performance improvement across all sites. Notably, the Dice score on Site E (out-of-federation) increases significantly from 28.23\% to 76.78\%. This gain highlights the effectiveness of temporal consistency in modeling surgical videos, where instrument appearance and context often persist across adjacent frames.}
Finally, our full FedST model (7th row) peaks the best performance across all sites, including external site E (76.99\%), verifying that different proposed components can complement each other to attain better results.

\subsubsection{Impact of Different Personalized FL settings}
To evaluate proposed FL setting is superior to other FL personalized settings, we set various personalized strategies for comparison in Table \ref{fl_setting}. \textbf{1st row}: all parameters are shared in FL training. \textbf{2nd row}: all parameters are personalized. \textbf{3rd row}: query embedding layers in the decoder are personalized. \textbf{4th row}: query embedding layers in both encoder and decoder are personalized. \textbf{5th row}: query embedding layers in the encoder \(\rho\)  which discussed in Sec. \ref{SERQ} are personalized. In Table \ref{fl_setting}, \textit{Attn.q} denotes query embedding layers in the attention-based module and \textit{Other} denotes all parameters in the decoder except query embedding layers.

It can be found that, when comparing the rows, the 2nd row only achieves acceptable performance on sites A-D, and does not work well on site E. This indicates that lacking parameter sharing leads to a lack of knowledge sharing, resulting in a loss of robustness and generalization ability in the global model. From the 3rd and 4th row, the privatization of the decoder parameters would actually harm the model's performance because the privatization of the encoder already separated the background tissue representation and consistent representation through the privatization of \(\rho\). Privatizing the decoder would lead to excessive separation which would weaken the consistent representation and hinder the ability of other sites to enhance it. On the other hand, the 5th row achieves the best average performance (82.69\%). This also suggests that keeping the parameter \(\rho\) is enough to separate the background representation and consistent instrument representation.

\subsubsection{Impact of Different Personalized Layers in PVT Backbone}
\begin{table}[t]
\centering
\renewcommand{\arraystretch}{1}
\setlength\tabcolsep{4pt}
\caption{\yj{Different Personalization Layers in Backbone}}
\resizebox{1\columnwidth}{!}{
\begin{tabular}{c|cccccc}
\hline\hline
\multirow{2}{*}{\centering Method} & \multicolumn{6}{c}{\textbf{Dice}(\%)$\uparrow$} \\
                                   & A & B & C & D & Avg. & E (Out Fed) \\
\hline
Norm           & 84.97 & 82.23 & 89.28 & 71.48 & 81.99 & 74.23 \\
Key            & 85.17 & 82.55 & \textbf{89.30} & 70.39 & 81.85 & 74.79 \\
Value          & \textbf{85.30} & \textbf{83.67} & 88.62 & 68.25 & 81.46 & 74.52 \\
Key\&Value     & 85.12 & 83.66 & 88.42 & 68.91 & 81.53 & 73.89 \\
Ours(Query)    & 84.73 & 83.15 & 89.13 & \textbf{71.46} & \textbf{82.62} & \textbf{77.97} \\
\hline\hline
\end{tabular}
}
\label{pvtq}
\end{table}

\begin{table}[t]
\centering
\renewcommand{\arraystretch}{1}
\setlength\tabcolsep{3pt}
\caption{\yj{Different Indicator in Channel Selection}}
\resizebox{1\columnwidth}{!}{
\begin{tabular}{c|cccccc}
\hline\hline
\multirow{2}{*}{\centering Method} & \multicolumn{6}{c}{\textbf{Dice}(\%)$\uparrow$} \\
                                   & A & B & C & D & Avg. & E (Out Fed) \\
\hline
Random   & 84.04 & 80.88 & 89.03 & 69.35 & 80.83 & 73.28 \\
Gaussian & \textbf{84.81} & 82.80 & 88.99 & 67.99 & 81.14 & 73.71 \\
One-hot  & 84.40 & 82.78 & \textbf{89.16} & 68.97 & 81.32 & 73.43 \\
Ours(CLIP)     & 84.73 & \textbf{83.15} & 89.13 & \textbf{71.46} & \textbf{82.62} & \textbf{77.97} \\
\hline\hline
\end{tabular}
}
\label{onehot}
\end{table}

\yj{To evaluate the effectiveness of different parameter personalization strategies within the PVT backbone, we compare various site-specific adaptations in Table~\ref{pvtq}. \textbf{1st row}: only the LayerNorm layers are personalized, while all other parameters are shared. \textbf{2nd row}: key embedding layers within the self-attention modules are personalized. \textbf{3rd row}: value embedding layers are personalized. \textbf{4th row}: both key and value embedding layers are personalized. \textbf{5th row}: our proposed setting, where only the query embedding layers in the PVT encoder are personalized.}

\yj{When only normalization layers are personalized (1st row), the effect is limited since normalization adjusts feature distributions but does not impact attention weights or semantic modeling, which are essential in transformer-based backbones.}
\yj{To better understand the effect of personalizing different components in attention, we refer to the self-attention formulation, where query ($Q$), key ($K$), and value ($V$) play distinct roles: queries define what each token wants to attend to, keys determine the comparison space, and values provide the semantic content to be aggregated.}
\yj{When only the key embeddings are personalized (2nd row), the model achieves the third-best performance. This can be attributed to the fact that keys, together with queries, determine the attention distribution. Due to the symmetry of the attention score computation, personalizing keys can have a similar effect to personalizing queries. However, keys are generally regarded as shared reference anchors across samples, and modifying them may disrupt the alignment of attention patterns across sites.}
\yj{When personalizing only the value embeddings (3rd row), the performance degrades. This is expected, as values represent the semantic output of the attention mechanism. Making them site-specific limits the model's ability to maintain a coherent feature space across sites, thereby reducing the benefits of federated collaboration.}
\yj{Personalizing both keys and values (4th row) further deteriorates performance. In this setting, both the attention computation and the aggregated semantics become site-specific, leading to a significant drop in generalization ability, particularly in the out-of-federation setting. The lack of shared structure hinders effective knowledge transfer and weakens the model’s global representation capacity.}
\yj{By personalizing only the queries, our method enables each site to adapt how attention is applied to its data while still benefiting from shared comparison structures (keys) and shared semantic content (values). This design is particularly suitable for surgical videos, where local spatial focus and instrument motion may vary across hospitals, but the overall semantic structure remains consistent.}
\begin{figure}[]
    \centering
    \includegraphics[width=\linewidth]{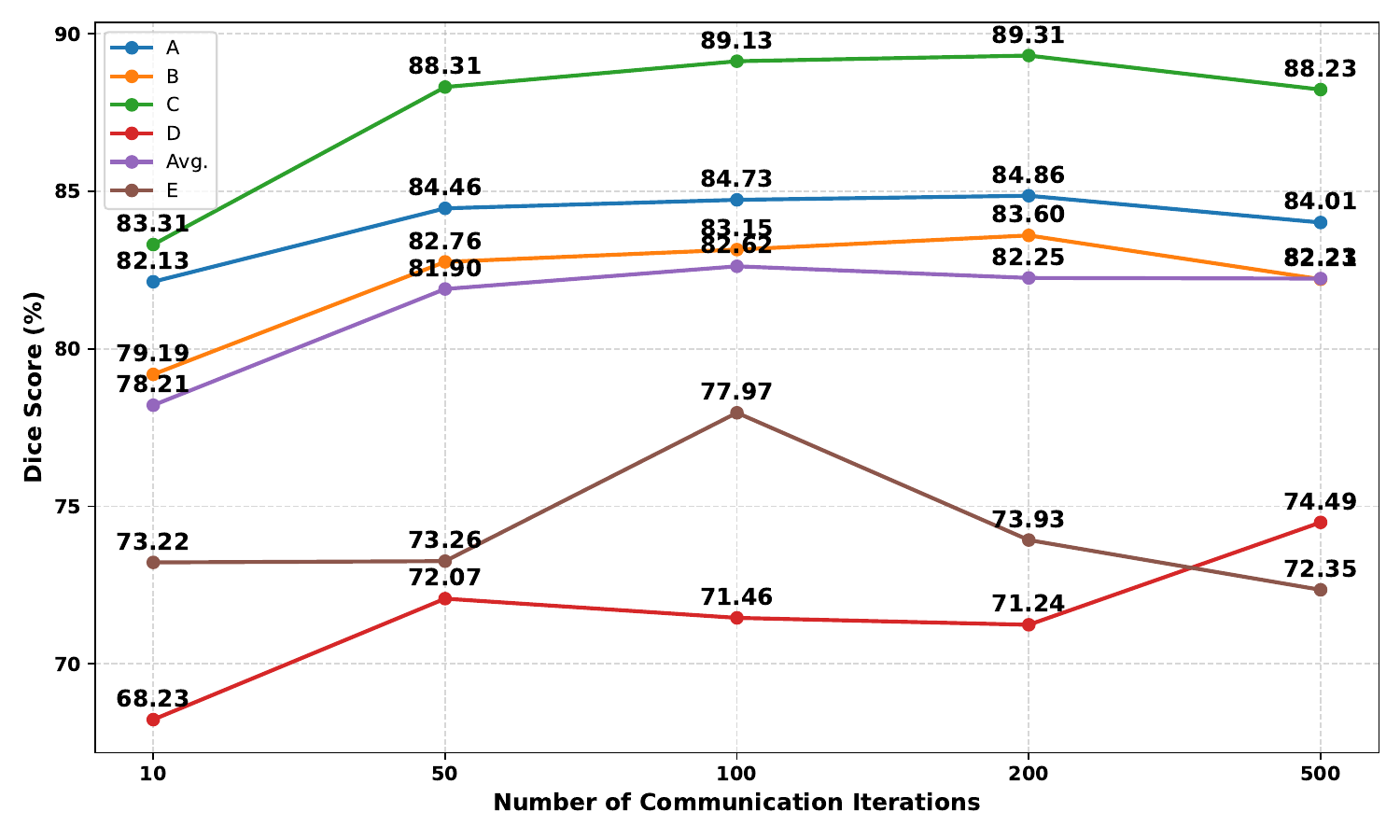}
    \caption{Effect of communication frequency on segmentation performance across sites.}
    \label{fig:comm_freq}
\end{figure}

\subsubsection{Impact of Different Indicators in Channel Selection Network}

\yj{To further investigate the effectiveness of our CLIP-based indicator, we compare it with several alternative designs in the channel selection network in Table \ref{onehot}. The detailed experimental settings for each method are as follows:
\textbf{1st row}: Randomly generates an embedding vector for each site without any semantic constraint.
\textbf{2nd row}: Generates embeddings using a Gaussian distribution centered at the one-hot value of each site. Specifically, the mean $\mu$ is determined by the site ID , and the standard deviation $\sigma$ is set to 0.1. This approach introduces slight variations around the one-hot indicator while maintaining a loose association with the site identity. The embedding is calculated as:  
$\exp\left(-\frac{(x - \mu)^2}{2 \cdot \sigma^2}\right)$
where $x$ represents the embedding dimensions.
\textbf{3rd row}: Uses a fixed one-hot encoding to represent each site. This setting assumes complete independence and orthogonality among sites.
\textbf{4th row}: Utilizes the CLIP-based indicator to capture the semantic relationships among different sites. This enables more informed and meaningful channel selection based on rich prior knowledge learned from CLIP.}

\yj{The results in Table~\ref{onehot} demonstrate clear performance differences among the evaluated methods. The Random indicator fails to provide any meaningful site-specific information, leading to poor performance due to the lack of explicit guidance for channel selection (e.g., 80.83\% average Dice and only 73.28\% on unseen Site E). The Gaussian indicator introduces slight variations around one-hot vectors but relies on artificial site ID mappings that do not capture real semantic relationships, resulting in limited improvements (average Dice 81.14\% and 73.71\% on Site E). The One-hot indicator performs slightly better by explicitly differentiating sites but still assumes strict independence and orthogonality, which prevents it from modeling shared characteristics across sites. This limitation is particularly evident on unseen Site E, where the performance remains low at 73.43\%. In contrast, our CLIP-based indicator effectively captures rich semantic relationships among sites by leveraging prior surgical domain knowledge, leading to consistently better results across all sites.}

\subsubsection{Effectiveness of SERQ in Using Synthetic Dataset} To validate the superiority of our FedST in leveraging synthetic data, we compare our method with other strategies for using synthetic data. The results are illustrated in Fig. \ref{vis_6}: the blue line represents that the synthetic data is directly utilized to fine-tune the averaged global model on the server side. The green line means that the synthetic data is utilized in each local site, to fine-tune the model for local optimization. Our method using SERQ is indicated by the red line.

We can see that our SERQ strategy can attain better results of all sites compared with others, especially for the site D with relatively limited data, demonstrating its effectiveness in leveraging the synthetic data by reducing the domain gap between synthetic and real surgical videos. Moreover, our method achieves the best model convergence speed and the most stable optimization process. This indicates that SERQ can successfully separate the domain representation in the synthetic dataset and leverage the instrument representation within the synthetic data to enhance the global model.

\subsubsection{Comparison of State-of-arts Methods Exploiting Synthetic Data for Training}
\begin{figure*}
    \centering
    \includegraphics[width=\textwidth]{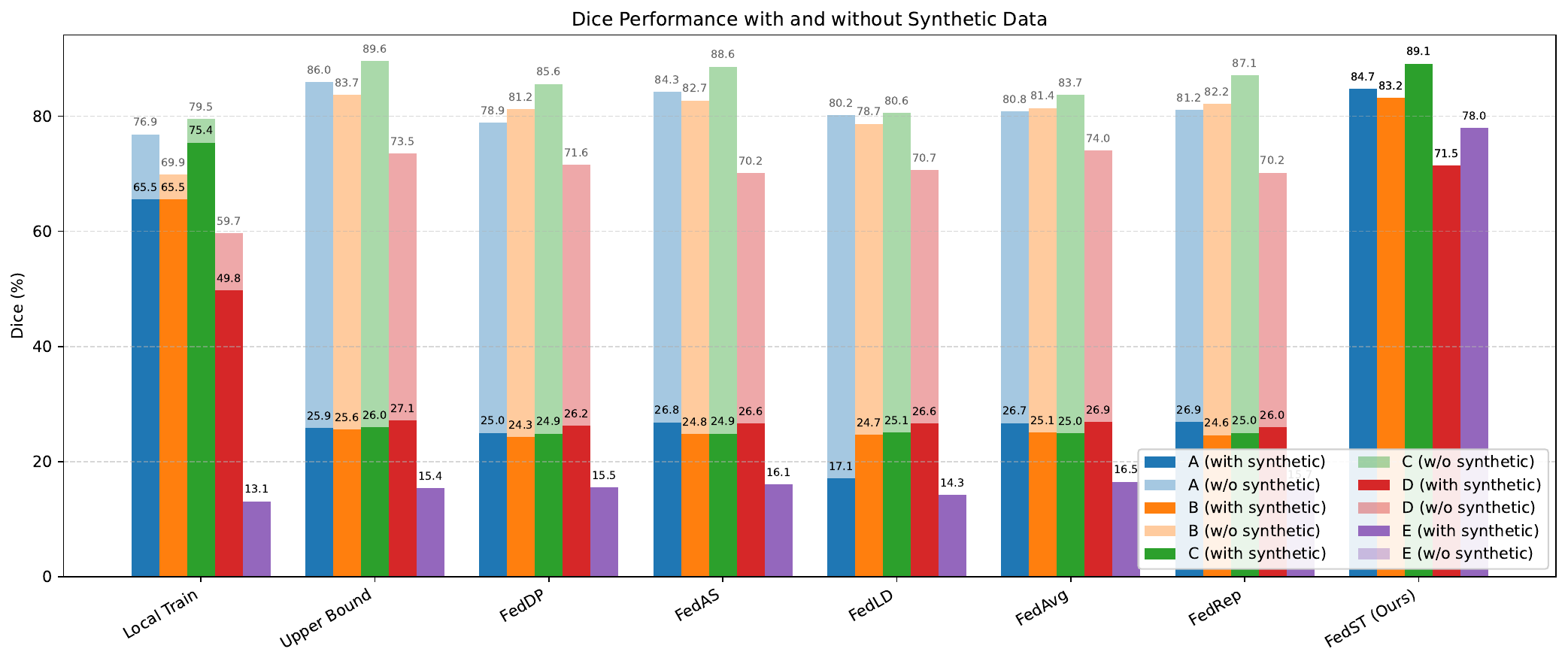}
    \caption{\yj{Dice Performance with and without Synthetic Data across All Sites.}}
    \label{synw}
\end{figure*}

\yj{In Figure~\ref{synw}, the lighter bars denote results without synthetic data, while the darker bars represent performance when synthetic data are used. In the synthetic setting, all synthetic samples are fully mixed into the real training data for each site. To ensure a fair comparison, the number of training iterations is doubled, so that the amount of real data exposure remains consistent with the original setting.} 

\yj{This visualization clearly reveals that, for baseline methods, using synthetic data actually degrades performance, especially on the in-federation sites A–D. This degradation can be attributed to two key factors. First, the domain gap between synthetic and real data causes models to overfit to unrealistic patterns or fail to converge. Second, in the federated setting, global model updates are aggregated across all sites. When each local model is trained on a mixture of real and synthetic data with mismatched distributions, the averaging process amplifies optimization conflicts, ultimately harming generalization. Even centralized training (Upper Bound) fails to converge under this setting.}
\yj{In contrast, FedST effectively leverages the synthetic data by introducing SERQ, a module designed to extract and enhance instrument representations. This enhancement enables stable training and improved generalization, as reflected in the consistently high performance of FedST across all sites.}

\subsubsection{Impact of Communication Frequency}

\yj{To evaluate how communication frequency affects federated training, we conduct an ablation study using different synchronization intervals: every 10, 50, 100, 200, and 500 local iterations. As shown in Figure~\ref{fig:comm_freq}, a moderate frequency of communication (e.g., every 100 iterations) results in the best segmentation performance across most sites. Frequent synchronization (e.g., every 10 iterations) slows down local convergence and introduces unnecessary overhead, while infrequent communication (e.g., every 500 iterations) leads to stale global updates and performance degradation. This analysis confirms the effectiveness of our chosen setting for efficient FL deployment.}

\subsection{Extensive Analysis}
\yj{In this part, we further study the generalization ability on unseen binary segmentation sites, as well as the computation and transmission efficiency of FedST.}
\subsubsection{Generalization on Unseen Binary Segmentation Sites}

\begin{table}[t]
\centering
\caption{\yj{Performance Comparison on Unseen Binary Segmentation GraSP Dataset}}
\renewcommand{\arraystretch}{1}
\setlength\tabcolsep{3pt}
\resizebox{1\columnwidth}{!}{
\begin{tabular}{l|cccc}
\hline\hline
Method & \textbf{Dice(\%)$\uparrow$}  & \textbf{IOU(\%)$\uparrow$}  & \textbf{HD95(\textit{pix.})$\downarrow$} & \textbf{ASSD(\textit{pix.})$\downarrow$}  \\ 
\hline
Local Train & 21.03  & 11.75  & 78.14  & 29.16 \\ 
Upper Bound & 26.46  & 15.25  & 73.23  & 27.23 \\
\hline
FedDP \cite{feddp} & 26.06  & 14.98  & 73.15  & 27.12 \\
FedAS \cite{yang2024fedas} & 26.40  & 15.21  & 74.05  & 27.49 \\
FedLD \cite{zeng2024tackling} & 26.26  & 15.12  & 73.21  & 27.30 \\
FedAvg \cite{fedavg} & 26.50  & 15.27  & 73.51  & 27.23 \\
FedRep \cite{fedrep} & 26.47  & 15.25  & 73.37  & 27.18 \\
FedST (Ours) & \textbf{90.64}  & \textbf{82.88}  & \textbf{21.70}  & \textbf{3.53} \\
\hline\hline
\end{tabular}
\label{grasp}
}
\end{table}

\yj{Due to the scarcity of surgical data and the high cost of manual annotation, the existing multi-class part segmentation datasets (Site A, B, and C) already represent the most accessible public resources for federated surgical segmentation. To further evaluate the generalization capability of our method beyond known segmentation settings, we conduct additional experiments on a binary segmentation task using the unseen GraSP\cite{grasp} dataset.}
\yj{The GraSP dataset contains 13 surgical videos and 3,490 video frames with binary annotations of surgical instruments. Compared to part segmentation, binary segmentation is a coarser-grained subtask, where all instrument parts are merged into a single foreground class. A reliable part segmentation model should, in principle, perform binary segmentation by simply treating all predicted parts as foreground. Therefore, strong performance on binary segmentation serves as an essential sanity check for evaluating generalization.}

\yj{As shown in Table~\ref{grasp}, Although binary segmentation is conceptually simpler than part segmentation, prior methods including FedAvg~\cite{fedavg}, FedRep~\cite{fedrep}, FedDP~\cite{feddp}, FedLD~\cite{zeng2024tackling}, and FedAS~\cite{yang2024fedas} struggle to generalize, achieving only around 26\% Dice. These results highlight the limitations of conventional FL methods when applied to unseen dataset.}
\yj{In contrast, our method FedST demonstrates remarkable generalization, achieving a Dice score of 90.64\% without any adaption on Grasp dataset. These results again highlight the importance of temporal modeling and instrument-specific representation enhancement in surgical video analysis.}

\subsubsection{Efficiency Analysis}
\begin{table}[t]
\centering
\caption{\yj{Communication and Efficiency Comparison across Methods}}
\renewcommand{\arraystretch}{1}
\setlength\tabcolsep{3pt}
\resizebox{1\columnwidth}{!}{
\begin{tabular}{l|ccc}
\hline\hline
Method & \textbf{Cost (MB)} $\downarrow$ & \textbf{Training Time (h)} $\downarrow$ & \textbf{FPS} $\uparrow$ \\
\hline
Local Train & 0.00   & $\sim$1.1 h    & $\sim$33.01 \\
Upper Bound & 0.00  & $\sim$32 h  & $\sim$33.01 \\
\hline
FedDP~\cite{feddp} & 6.7  & $\sim$8.3 h         & $\sim$33.01 \\
FedAS~\cite{yang2024fedas} & 19.71  & $\sim$9.6 h         & $\sim$33.01 \\
FedLD~\cite{zeng2024tackling} & 19.71  & $\sim$10.9 h         & $\sim$33.01 \\
FedAvg~\cite{fedavg} & 19.71  & $\sim$8.3 h  & $\sim$33.01 \\
FedRep~\cite{fedrep} & 13.01  & $\sim$8.3 h  & $\sim$33.01 \\
\hline
\textbf{FedST (Ours)} & 12.69 & $\sim$20 + 23.8 h & $\sim$23.04 \\
\hline\hline
\end{tabular}
\label{tab:comm_efficiency}
}
\end{table}

\yj{We compare the communication and efficiency characteristics of our method with several representative FL baselines, as summarized in Table~\ref{tab:comm_efficiency}. The table reports three key metrics: (1) communication cost per round, calculated based on the number of transmitted trainable parameters; (2) training time, including both local and server-side computation; and (3) FPS, measuring the model's inference speed on real surgical video frames.}

\yj{FedST achieves a notably lower communication cost (12.69 MB) compared to fully shared methods such as FedAvg and FedLD (19.71 MB). This efficiency results from the partial personalization of the RSC module—since its parameters are site-specific and not communicated, the overall transmission size is significantly reduced.}
\yj{In terms of training time, FedST requires $\sim$43.8 hours in total, consisting of $\sim$20 hours of local training and $\sim$23.8 hours of server-side computation. In contrast, most baselines involve negligible server-side computation, as their training is entirely performed on local sites. The increased local training time in FedST is due to the introduction of temporal modeling through the RSC module. Additionally, the server-side SERQ module requires extra computation time because it utilizes synthetic data to enhance global representations.}
\yj{It is important to note that during inference, only the RSC module is activated on the local site side. In contrast, SERQ operates solely during training and is never involved in inference. Although temporal modeling via RSC slightly reduces FPS (from $\sim$33.01 to $\sim$23.04), our method still supports real-time inference, demonstrating its practical applicability in surgical video analysis.}

\section{Discussion}

RAS presents unique characteristics in FL, including background diversity with instrument similarity and large-scale open-source synthetic data with huge domain gap. Previous FL methods often combine background and instrument features, failing to distinguish their distinct roles in segmentation. While PFL approaches attempt to decouple certain parameters to personalize the background, they neglect the explicit enhancement of instrument-consistent representation. This oversight leads to inconsistent convergence across models and poor performance. 

In this paper, we propose \textbf{FedST} to address these limitations by introducing RSC and SERQ. 
Our approach explicitly separates and enhances instrument-consistent representations, leveraging temporal information and a lightweight network with pre-trained CLIP. We construct temporal representations and selectively enhance the channels with site-level knowledge to improve model performance.  
% Although previous methods, While FedDP~\cite{feddp} also explores parameter personalization by modifying query embeddings in self-attention, its focus lies in capturing pixel-level dependencies within static medical images for segmentation. In contrast, our method targets spatiotemporal consistency and domain-specific tissue background modeling across surgical video clips.
\yj{Although previous methods have explored different forms of personalization, FedDP~\cite{feddp} focuses on modifying query embeddings in self-attention to capture pixel-level dependencies within static medical images. FedRep~\cite{fedrep}, on the other hand, personalizes the decoder layers at the prediction level to adapt to site-specific outputs. In contrast, our method operates at the representation level and targets spatiotemporal consistency and domain-specific tissue background modeling across surgical video clips, enabling robust generalization under inter-site variability.}
% 
% We initially explore to use one-hot embeddings to guide channel selection. While this approach provided modest performance gains, it lacks interpretability and treats each site as orthogonal, ignoring potential relationships between sites.
% %
% Previous methods have explored leveraging synthetic datasets to address inconsistent convergence of the global model. \cite{zhu2021data, lin2020ensemble} utilized synthetic datasets to enhance the global model but overlooked the impact of domain gaps between synthetic and real datasets.
% % \cite{lin2020ensemble} and 
% %
% Our SERQ strategy overcomes these limitations by employing a moving average technique to iteratively update the pre-trained model, explicitly quantifying the instrument-consistent representation. This representation is then used to enhance the global model to synchronize the convergence of local models.  

Remarkably, even without additional training on the out-of-federation site, the global model performs exceptionally well when directly applied for inference on this site. This impressive result underscores the effectiveness of our model in capturing similar temporal representations and consistent instrument representations across diverse sites. Thus, our approach is not limited to surgical scenarios and can be applied to any domain where similar temporal information exists across different sites.
During inference, our method only requires the RSC module, which has significantly lower computational cost compared to the SERQ module. As a result, our method can achieve rapid inference speed with 23.06 FPS. Furthermore, substituting the encoder with more efficient architectures could potentially enhance inference speed even further. While the FedST framework has shown significant improvements in addressing background diversity and domain gap in RAS, it has certain limitations. One key challenge is its adaptability to other surgical tasks, such as workflow recognition and depth estimation. Future work will focus on extending FedST to a multi-task framework.
\yj{In addition, although FedST demonstrates strong generalization across diverse surgical sites, it still struggles under challenging visual conditions. Specifically, we observe that the model's performance degrades in frames with poor image quality, such as specular reflections, motion blur, and occlusions caused by tissue overlap. These factors reduce the visibility of surgical instruments, impairing both representation learning and segmentation accuracy. Addressing these issues may require incorporating low-level enhancement modules or uncertainty-aware training strategies in future.}

\section{Conclusion}
This paper introduces a novel cross-type PFL framework, FedST, designed to address background diversity and the domain gap in synthetic surgical datasets for RAS, involving data from multiple surgery types. It begins by explicitly identifying a consistent instrument representation through RSC, using a motion model and textual guided channel selection to guide each site in focusing on the instrument during FL. Subsequently, SERQ defines an explicit representation target based on synthetic data, synchronizing model convergence during fusion.
Our method has been validated across five surgical sites from five independent public datasets and consistently achieves superior results over state-of-the-art FL methods on all sites, especially the out-of-federation site.
\vspace{0.5cm}
\input{} % <- 这里是空的会编译错误，如非占位请删掉或填文件名

\bibliographystyle{IEEEtran}
\bibliography{LaTeX/mybib}

\end{document}